\documentclass[aps,prd,preprint,superscriptaddress,preprintnumbers,nofootinbib]{revtex4-1}
\pdfoutput=1

\usepackage{graphics}

\usepackage[dvips]{graphicx}
\usepackage{mathrsfs}
\usepackage{amssymb}
\usepackage{amsmath}
\usepackage{verbatim}
\usepackage{float}
\usepackage{slashed}
\usepackage{bbm}
\usepackage[dvips,letterpaper,text={6.5in,9in}]{geometry}

\usepackage[english]{babel}
\usepackage{tabu, multirow,color,dcolumn,bm,enumerate}

\newcommand{\tev}{\text{TeV}}
\newcommand{\gev}{\text{GeV}}
\newcommand{\mev}{\text{MeV}}
\newcommand{\fb}{\text{fb}}

\usepackage{graphicx,bm}

\newcommand{\MET}{$E_T{\hspace{-0.47cm}/}\hspace{0.35cm}$}
\newcommand{\met}{\slashed{E}_T}

\newcommand{\Lc}{\mathcal{L}}

\newcommand{\Mc}{\mathcal{M}}

\newcommand{\beq}{\begin{equation}}
\newcommand{\bea}{\begin{eqnarray}}
\newcommand{\eeq}{\end{equation}}
\newcommand{\eea}{\end{eqnarray}}
\newcommand{\bal}{\begin{align}}
\newcommand{\eal}{\end{align}}

\usepackage{tikz}
\usetikzlibrary{decorations.pathmorphing,decorations.markings}
\tikzset{
photon/.style={decorate, decoration={snake,amplitude=4pt, segment length=7pt}, draw=black},
particle/.style={draw=black, postaction={decorate}, decoration={markings,mark=at position .5 with {\arrow[draw=black]{>}}}},
antiparticle/.style={draw=black, postaction={decorate}, decoration={markings,mark=at position .5 with {\arrow[draw=black]{<}}}},
gluon/.style={decorate, draw=black, decoration={coil,amplitude=3pt, segment length=4pt}},
higgs/.style={draw=black,dashed,thick },
arrow/.style={draw=black, very thick, postaction={decorate}, decoration={markings,mark=at position 1 with {\arrow[draw=black]{>}}}}
}

\usepackage{latexsym}
\usepackage{subcaption}

\usepackage{outlines}
\usepackage{enumitem}
\setenumerate[1]{label=\Roman*.}
\setenumerate[2]{label=\Alph*.}
\setenumerate[3]{label=\roman*.}
\setenumerate[4]{label=\alph*.}

\definecolor{darklightsabergreen}{rgb}{0.0, .49, 0.06}

\begin{document} 

\title{ Using the (Modified) Matrix Element Method to constrain $L_\mu - L_\tau$ Interactions }

\author{Fatemeh Elahi}
\affiliation{Department of Physics, 225 Nieuwland Science Hall, University of Notre Dame, Notre Dame, IN 46556, USA}
\author{Adam Martin}
\affiliation{Department of Physics, 225 Nieuwland Science Hall, University of Notre Dame, Notre Dame, IN 46556, USA}

\vspace*{0.5cm}
\begin{abstract}
\vspace*{0.5cm}

{
In this paper, we explore the discriminatory power of the matrix element method (MEM) in constraining the $L_\mu-L_\tau$ model at the LHC. The $Z'$ boson associated with the spontaneously broken $U(1)_{L_\mu-L_\tau}$ symmetry only interacts with the second and third generation of leptons at tree level, and is thus difficult to produce at the LHC. We argue that the best channels for discovering this $Z'$ are in $Z \to 4\mu$ and $2\mu+\met$. Both these channels have a large number of kinematic observables, which strongly motivates the usage of a multivariate technique. The MEM is a multivariate analysis that uses the squared matrix element $|\Mc|^2$ to quantify the likelihood of the testing hypotheses. As the computation of the $|\Mc|^2$ requires knowing the initial and final state momenta and the model parameters, it is not commonly used in new physics searches. Conventionally, new parameters are estimated by maximizing the likelihood of the signal with respect to the background, and we outline scenarios in which this procedure is (in)effective. We illustrate that the new parameters can also be estimated by studying the $|\Mc|^2$ distributions, and, even if our parameter estimation is off, we can gain better sensitivity than cut-and-count methods. Additionally, unlike the conventional MEM, where one integrates over all unknown momenta in processes with \MET, we show an example scenario where these momenta can be estimated using the process topology. This procedure, which we refer to as the ``modified squared matrix element", is computationally much faster than the canonical matrix element method and maintains signal-background discrimination. Bringing the MEM and the aforementioned modifications to bear on the $L_\mu-L_\tau$ model, we find that with $300 \, \fb^{-1}$ of integrated luminosity, we are sensitive to the couplings of $ g_{Z'} \gtrsim 0.002 ~ g_1$ and $M_{Z'} < 20 ~ \gev$, and $g_{Z'} \gtrsim 0.005 ~ g_1$ and $20\, \gev <M_{Z'} < 40 ~ \gev$, which is about an order of magnitude improvement over the cut-and-count method for the same amount of data. 
} 
\end{abstract}
\maketitle

\section{ Introduction}
\label{sec:Intro}

The highest priority of LHC-- run II is to find physics beyond the standard model (BSM). In the pursuit of optimal sensitivity to BSM physics, it is important to develop a diverse set of techniques that improve our sensitivity to BSM signal against the SM background. The most straightforward technique is the cut-and-count method, where one determines the suitable region of phase space by studying the kinematic distributions of the simulated events (e.g. the invariant mass of two detected particles, or the angle difference between them) and look for an excess over the SM expectation. While simple, this method can miss correlations among kinematic variables. To improve our discriminatory power, one can select events based on more complicated variables that take into account correlations. These complicated variables can be a linear or a non-linear combination of the kinematic variables used in the cut-count method. The techniques that consider combinations of kinematic variables are generally called ``Multivariate Analysis" (MVA)~\cite{Hocker:2007ht}. Some of the main methods developed in this category are ``Neural Network"~\cite{Denby:1987rk,Bellantoni:1991ei, Bortoletto:1991ty, Babbage:1993ry, Becks:1993jf} and ``Boosted Decision Tree"~\cite{Roe:2004na,Yang:2005nz}. However, these MVA methods usually require a phase of computer training and techniques that are not transparent to theorists. Another example of MVA that is calculated based on the theoretical assumptions for a given process is the Matrix Element Method (MEM)~\cite{Fiedler:2010sg,Artoisenet:2008zz,Mattelaer:2011ywa,Mertens:2014iya,Bianchini:2016yrt,Volobouev:2011vb,Biedermann:2016lvg}.

The MEM was originally developed in the Tevatron experiment and was successfully applied to measurement of the top quark mass and single top electroweak production~\cite{Abazov:2004cs, Abazov:2005zz, Kondo:1991dw, Dalitz:1991wa, Abulencia:2007br, Aaltonen:2008bd, Abazov:2008ds, Abazov:2009ii, Aaltonen:2010jr, Fiedler:2010sg, D0:2016ull, Bianchini:2016yrt, Cortiana:2015rca,Schouten:2014yza,Gainer:2013iya,Abazov:2011fc}. In the MEM, given a reconstructed event and a theoretical hypothesis, the probability that the event matches the hypothesis is quantified based on the value of the partonic matrix element for the hypothesis processes evaluated using the final state four momenta. Repeating this procedure using multiple hypothesis processes, e.g. a background process and a signal process, and comparing the results, one can quantify how `signal-like' or `background-like' a given ensemble of events is. By definition, the MEM contains all of the kinematic information of the hypothesis process so it captures all correlations. Moreover, the MEM has a clear physical meaning and there is a transparent link between the theoretical assumptions and event reconstruction.

Despite the successes of the MEM in the SM measurements, this method has not been applied extensively in BSM searches. Some of the main difficulties of the MEM are the following~\cite{Fiedler:2010sg, Volobouev:2011vb}:
 \begin{itemize}
 \item The squared matrix element depends on all of the momenta -- both initial and final -- in the event. In circumstances where one or more momenta is not determined uniquely, we must integrate over all possible values. The more integrations required for each event, the more time and computational resources required. Events with missing energy, a hallmark of many new physics scenarios, automatically fall into this category of events with unconstrained momenta. 
 
 \item BSM hypotheses necessarily introduce new parameters ($\alpha = $ masses, couplings of new particles), which need to be determined before we can calculate the matrix element. Without a separate experiment telling us what $\alpha$ to use, our best option is to choose the $\alpha$ such that they maximize our reach for detecting the signal given the background. However, finding the best-suited values for these parameters can be difficult, as shown in Section \ref{sec:fourmu}. 
 \end{itemize}

 Ideally, we would like to develop techniques to overcome the aforementioned challenges for \emph{any} BSM hypothesis. However, as a first step, we focus on dealing with some of these challenging for a specific model. The BSM hypothesis we will consider in this paper is one of the simplest extensions of the SM, the $L_\mu-L_\tau$ model~\cite{He:1991qd}. In the $L_\mu-L_\tau$ model, the difference between muon number and tau number -- an accidental symmetry of the SM -- is gauged and spontaneously broken, generating a massive $Z'$ that interacts only with the second and third generation of leptons at tree level~\cite{He:1991qd}. As the $L_\mu- L_\tau\, Z'$ is hadrophobic and does not interact with electrons at tree level, it is difficult to produce at conventional experiments and therefore is challenging to constrain. In fact, tree-level $Z'$ production at the LHC always involves four second/third generation lepton (either charged or neutral) final states: the initial partons create a pair of leptons via $W^{\pm}/Z/\gamma$ exchange, one of which radiates a $Z'$ that subsequently decays to a second lepton pair. At the LHC, identifying muons is much easier and more accurate than identifying taus, thus in this work, we will only consider combinations of muons and \MET (due to $\nu_\mu$ or $\nu_\tau$) as potential final states. In particular, we will study the two processes of i) four muons ($4\mu$), and ii) two muons and missing energy ($2\mu + \slashed E_T$). Of course, as all of the final state particles are SM particles, there will be interference between the four muon events produced via $Z'$ and SM four muon events. We will put this interference to use in sections \ref{sec:fourmu} and \ref{sec:twomu}.

The four muon final state is fully reconstructable, and it has 12 independent observables. The existence of this many kinematic variables begs for using an MVA. The MEM is a suitable choice because it optimally~\cite{Fiedler:2010sg} uses all of the available kinematic variables to distinguish signal from the background. Applying MEM to the process $2 \mu+ \met$ with 6 observables is also more lucrative, however, the existence of \MET makes its computation more challenging. Fortunately, the main SM background to this process ($p p \to \tau \tau \to 2 \mu\, \met$) has a very specific topology, and thus the missing momenta for this process can be estimated from the momenta of muons. In section \ref{sec:twomu}, we show that by using the squared matrix element of the $ \left.\tau^+ \tau^- \right|_{\text{dimuon}}$ background as a discriminatory variable (as opposed to the conventional likelihood function) with the guessed missing momenta derived from its topology (i.e. without having to integrate over missing momenta), we are able to sufficiently differentiate our signal from the remaining SM backgrounds as well. Due to the deviation from the canonical MEM procedure, we will refer to this approach as the ``modified MEM".

 The organization of the rest of the paper is as follows. In the next section, we explain the Matrix Element method and how it is used to discriminate the signal from the background. In section \ref{sec:model}, we introduce the model and discuss some of the constraints on its parameters from LHC and other experiments. In section \ref{sec:analysis}, we explore how the bounds can be improved at the LHC. Finally, a discussion about the results and concluding remarks are made in section \ref{sec:discuss}.

\section{ Matrix Element Method}
\label{sec:MEM}

 At the LHC, we are searching for BSM physics against the SM background. Although we could look for generic departure from the SM, our sensitivity is greater if we look for a particular BSM physics hypothesis. Therefore, we are usually dealing with two hypotheses: a specific new physics model (signal) and the null hypothesis (the SM background). To statistically analyze these two test hypotheses, the MEM uses their likelihood ratio, where the likelihoods are calculated based on the squared matrix element ($|\Mc|^2$) of a process at tree level\footnote{ Some papers have studied the MEM at NLO as well~\cite{Baumeister:2016maz, Martini:2015zkx, Martini:2015fsa}.}, and evaluated using the final state momenta of individual events. Since all the kinematic information of a process is contained in the matrix element, $|\Mc|^2$ is a powerful variable to discriminate between the two test hypotheses. If the empirical LHC events are inconsistent with either of the conjectures, $|\Mc|^2$ should also favor neither of them.

 The function $|\Mc|^2$, depends on the momentum of initial and final state partons ($p^{\rm par}$). The squared matrix element of the signal $|\Mc|^2_{\text{signal}}$ will additionally depend on the new model parameters ($\alpha$): $|\Mc|^2_{\text{signal}} = |\Mc ( p^{\rm par}; \alpha)|^2$, and $|\Mc|^2_{SM} = |\Mc ( p^{\rm par})|^2$. In the special case where the final states of the signal exactly match the SM background (e.g, no BSM particles in the final state), $|\Mc|^2_{\text{signal}}$ must include the interference term with the SM as well.
 
 If the LHC could detect the $p^{\rm par}$ of all of the final state particles, the likelihood $(P)$ that an observed event is due to a particular hypothesis would be defined as 

 \beq
 P(p^{\rm par} |\alpha ) = \frac{1}{\sigma} \int dx_{_1} dx_{_2} \frac{f(x_{_1})f(x_{_2})}{2 s x_{_1}x_{_2}} \ |\Mc ( p^{\text{par}};\alpha)|^2 \delta^4 (p_{\text{initial}}^{\text{par}} -p_{\text{final}}^{\text{par}}), 
 \label{eq:definition1}
 \eeq
where $x_i$ and $p_{\text{initial}}^{\text{par}}$ are intimately related: $p_{\text{initial,i}}^{\text{par}} \equiv \frac{\sqrt{s}}{2}(x_i, 0, 0,\pm x_i)$. The factors $f(x_i)$ are the parton distribution functions (PDF) of the initial states, the collider center of mass energy of collision is represented by $\sqrt{s}$, and $\sigma$ is the total cross section with which $P(p^{\rm par}|\alpha)$ is normalized to ensure $\int P (p ^{\text{par}}|\alpha ) dp^{\text{par}}= 1$. The factor $\delta^4 (p_{\text{initial}}^{\text{par}} -p_{\text{final}}^{\text{par}})$ ensures the conservation of energy and momentum in the process. If we know the final state four-momenta, we can use this delta function to infer information about the initial state momenta. More specifically, the delta function involving the $p_{\text{initial}}^{\text{par}}$ can be translated to a delta function on $x_i$'s, uniquely defining the $x_i$'s and collapsing the integrals: 

\begin{align}
\delta^4 (p_{\text{initial}}^{\text{par}} -p_{\text{final}}^{\text{par}})=& \delta(\frac{\sqrt{s}}{2} (x_{_1}+x_{_2})- p_{\text{final}}^{\text{par, Energy}})\times \nonumber\\
& \delta^2 ( p_{\text{final}}^{\text{par, transverse}})\times \nonumber\\
&\delta(\frac{\sqrt{s}}{2} (x_{_1}-x_{_2})- p_{\text{final}}^{\text{par, longitudinal}}). 
\label{eq:deltas}
\end{align}

 The events at the LHC, however, are defined according to reconstructed momenta ($p^{\rm rec}$) at the detector which may not equal the $p^{\rm par}$. Hence, we must modify the likelihood $(P)$ to be a function of detector level momenta: 
 
 \beq
 P(p^{\rm rec} |\alpha ) = \frac{1}{\sigma} \int d\Phi(p^{\text{par}}_{\text{final}}) dx_{_1} dx_{_2} \frac{f(x_{_1})f(x_{_2})}{2 s x_{_1}x_{_2}} \ |\Mc ( p^{\text{par}};\alpha)|^2 \delta^4 (p_{\text{initial}}^{\text{par}} -p_{\text{final}}^{\text{par}}) W(p^{\rm rec}, p^{\rm par}),
 \label{eq:definition}
 \eeq
 
 with an integration over the possible values of partonic momenta given the reconstructed momenta (represented by $\int d\Phi(p^{\rm par}_{\text{final}}) dx_{_1} dx_{_2}$). When there is missing energy (e.g, neutrino) in the event, we integrate over all unconstrained momenta. However, if the particle is detected, we use the transfer functions $W(p^{\rm rec}, p^{\rm par})$ to translate between the partonic momenta and reconstructed momenta. These functions are usually gaussian (or bi-gaussian), where the arguments are estimated according to Monte Carlo (MC) simulations\footnote{In the matrix element technique, the theoretical assumptions (the $|\Mc|^2$) and the assumptions about the experiment and detectors (the transfer functions) factorize and are independent of each other. Therefore, the improvement in any of these assumptions can easily be implemented in the matrix element method.}~\cite{Artoisenet:2008zz,Mattelaer:2011ywa,Mertens:2014iya}. If the detected particle is a lepton or a photon, the transfer function are well approximated by a delta function $(\delta(p^{\text{par}} - p^{\text{rec}})$), while for colored objects, the reconstructed momenta may be significantly different from the partonic momenta.

 After calculating the likelihood function ($P$) of the signal and the SM, we need to determine whether a given event is more likely to be due to the signal hypothesis or the background. Therefore, we study the likelihood ratio:
 
 \beq
 \Lc (p_i^{\rm rec};\alpha) = \frac{P( p_i^{\rm rec}| \text{new phyiscs } (\alpha))}{ P( p_i^{\rm rec}| \text{null hypothesis})},
 \label{eq:likelihoodratio}
 \eeq
 where the $i$ subscript refers to the $i$-th event. If the value of $\alpha$ were known, we could plot the distribution of $\Lc$ for given events, just like any usual kinematic distribution that uses the reconstructed momenta. In the $\Lc$ distribution, larger values would indicate the signal hypothesis is favored and lower values meant the data is more consistent with the null hypothesis. Hence, with a cut on the larger values of $\Lc$, we could find the phase space that increases the signal fraction.
 
 Having said that, in the case of new physics, we do not know the value of $\alpha$. Since we want to optimize our reach for the signal hypothesis, we want to choose $\alpha$ that maximizes $ \Lc (p_i^{\rm rec};\alpha)$ (or equivalently $log \left[ \Lc (\alpha)\right]$). Therefore, we plot $log \left[ \Lc (\alpha)\right]$ with respect to $\alpha$ and look for maxima in the plot. This process can be done for each event. For multiple events, we simply sum over $i$:
 \beq
 \sum_i \log [\Lc (p_i^{\rm rec};\alpha)] = \sum_i \log \left[ \frac{P( p_i^{\rm rec}| \text{new phyiscs } (\alpha))}{ P( p_i^{\rm rec}| \text{null hypothesis})}\right].
 \label{eq:plot}
 \eeq
 
 We denote $\alpha^*$ for the value of $\alpha$ that maximizes Eq.~(\ref{eq:plot}). As it will be important later on, we emphasize that Eq.~(\ref{eq:plot}) only yields one number for an entire set of events. Previous studies have shown that if $\alpha$ is the mass of a particle, the $\alpha^*$ returned by maximizing Eq.~(\ref{eq:plot}) is actually the same as the true value of $\alpha$; in fact, the current most precise measurement of the top quark mass is obtained with this process of maximizing the likelihood ratio with respect to top mass~\cite{Fiedler:2010sg,Bianchini:2016yrt}. This procedure is used only to determine $\alpha^*$, and it gives no information on the prospect of discovering the signal. Once the optimal value $\alpha^*$ is determined, the sensitivity to the signal is determined by selecting a region of the $\Lc (\alpha^*)$ (with $\alpha = \alpha^*$ in Eq.~(\ref{eq:likelihoodratio})) distribution that optimizes the signal over background ratio. 
 
 Despite the power of the MEM in discriminating signal against the null hypothesis, due to its computational difficulties, it is not commonly used in BSM searches. One reason may be the number of phase space integrations in the calculation of $P(p^{\rm rec} |\alpha )$. If the process of interest has multiple sources of missing energy or contains colored objects, the number of integrals can be high and may over-consume computational resources. Another reason is that the maximization of likelihood ratio with respect to $\alpha$ can be challenging depending on the nature of parameters. As we will see in Section 3.A, the maximization of log likelihood works if $\alpha$ is the mass of a new particle, but this approach breaks down if $\alpha$ is a coupling. That is because $log \left[ \Lc (\alpha)\right]$ only increases with respect to $\alpha$ and has no local maxima. Although one naively might expect that larger values of $\alpha$ results in greater sensitivity to BSM hypothesis, in section \ref{sec:fourmu}, we will show that is not correct.
 
 In this paper, we will study a simple model to give a working example of how some of the challenges in MEM for BSM can be overcome. The example we will consider is the gauged $U(1)_{L_\mu-L_\tau}$ symmetry, summarized in the next section. This model contains a massive $Z'$ gauge boson that only interacts with the second and third generation of leptons. We will show that the LHC sensitivity can be improved by an order of magnitude compared with the cut-and-count method if we apply MEM. 
 
\section{ $L_\mu-L_\tau$ model}
\label{sec:model}

The difference between muon number and tau number $L_\mu-L_\tau$ is one of the accidental global symmetries present in the SM. This $U(1)_{L_\mu - L_\tau}$ symmetry is anomaly free and therefore can be gauged. However, from the oscillation of tau or muon neutrinos to electron neutrinos, we know this symmetry is not respected in nature and needs to be broken. The consequence of the breaking is a neutral, color singlet, massive $Z'$ that couples only to muon number and tau number at tree level. The interactions of $Z'$ are described by the Lagrangian below:
\begin{align}
\mathcal L \ni- \frac{1}{4} (Z')_{\alpha \beta} (Z')^{\alpha \beta} + \frac{1}{2} M_{Z'}^2 Z^{'\alpha}Z'_{\alpha} - \epsilon g_1 Z'_{\alpha} \left( \bar \ell_2 \gamma^\alpha \ell_2 + \bar \mu \gamma^\alpha \mu - \bar \ell_3 \gamma^\alpha \ell_3 - \bar \tau \gamma^\alpha \tau\right),
\end{align}
where $Z'_{\alpha \beta} = \partial _\alpha Z'_\beta - \partial_\beta Z'_\alpha$ is the field strength tensor, and $\ell_2 = (\nu_\mu, \mu)^T$, $\ell_3 = (\nu_\tau, \tau)^T$. As shown in the Lagrangian, the $Z'$ has the same coupling to left handed and right-handed muon (tau), with a relative minus sign between the coupling of muons and taus~\cite{He:1991qd}. The new parameters in the model are $\alpha = (M_{Z'} , \epsilon)$, where $ \epsilon g_1$ is the coupling of $Z'$ to muons and taus. 

Studying this model is important because some region of its parameter space can explain the long persisting discrepancy in the SM prediction and experimental measurement of muon anomalous magnetic moment $(g-2)_\mu$~\cite{Anastasi:2015oea, Baek:2001kca,Ma:2001md,Gninenko:2001hx,Pospelov:2008zw,Heeck:2011wj,Harigaya:2013twa,Chen:2017cic}. Some anomalies observed in B physics and flavor changing Higgs coupling~\cite{Crivellin:2015lwa,Crivellin:2015mga,Heeck:2014qea} can also be explained by gauged $L_\mu-L_\tau$, which further motivates studying this model. In particular, the anomalies recently observed in $R_{K}= \text{Br} (B \to K \mu^+ \mu^-)/\text{Br} (B \to K e^+ e^-) \simeq 0.745 $ with $2.6\, \sigma$ discrepancy between theoretical expectations, and in $R_{K^{*}}= \text{Br} (B \to K^{*} \mu^+ \mu^-)/\text{Br} (B \to K^{*} e^+ e^-) \simeq 0.7 $ with $\sim 2.5\, \sigma$, by LHCb~\cite{Bifani:2260258} can be explained by the $L_\mu-L_\tau$ model, assuming $M_{Z'} \sim O( \tev) $ and $g_{Z'} \sim O(1)$~\cite{Crivellin:2015lwa,Crivellin:2015mga, Alonso:2017bff}. 

Because the $L_\mu-L_\tau \ Z'$ only interacts with second and third generation leptons at tree level, it is not very constrained. One constraint comes from $Z-Z'$ mixing that arises from loops of muons and taus (and their respective neutrinos), inducing a coupling of $O(10^{-3}\,\epsilon)$ between the $Z'$ and all fermions. The factor of $10^{-3}$ is a rough estimate based on the loop suppression and the couplings of $Z'$ and $Z$ with muon and tau. Precision measurements of $Z-$electron coupling at the BaBar and Belle II experiment~\cite{Eigen:2015rea,Curtin:2014cca,Eigen:2015sea,Essig:2013vha,Wang:2015hdf,Araki:2017wyg} requires the coupling of $Z'$ to electrons to be $ \lesssim 10^{-3}$ for $10 \ \mev < M_{Z'} < 10 \ \gev$, which translates to $ \epsilon \lesssim 1$ in the $L_\mu-L_\tau$ model. 

The strongest current bound on $L_\mu-L_\tau$ for $M_{Z'} < 10 \ \gev$ is from fixed target neutrino beam experiments. In particular, CHARMII~\cite{Geiregat:1990gz} and CCFR~\cite{Mishra:1991ws, Mishra:1991bv} tightly constrain $L_\mu-L_\tau \ Z'$ via the trident process: $N+ \nu_\mu \rightarrow N+ \nu_\mu + \mu\mu$~\cite{Altmannshofer:2014cfa}. This process occurs through the exchange of off-shell $W^\pm/Z$ bosons in the SM. However, in the $L_\mu-L_\tau$ model, the exchange of $Z'$ can significantly contribute to the rate of the process, especially if the $Z'$ is produced on-shell ~\cite{Geiregat:1990gz,Mishra:1991ws,Altmannshofer:2014cfa,Kaneta:2016uyt}. Neutrino trident experiment excludes a $L_\mu-L_\tau \ Z'$ with $ \epsilon \gtrsim 0.005 $ and $M_{Z'} \lesssim 1\ \gev$. The bounds loosen for heavier $Z'$ to $\epsilon \sim 0.05$ for $M_{Z'} = 20 \ \gev$.

LHC bounds on this model come from from recasting the $pp \to Z \to 4\mu$ searches by CMS and ATLAS~\cite{CMS:2012bw, Aad:2014wra}. This bound surpasses the trident bound for $10 \ \gev< M_{Z'} \lesssim 45 \, \gev$~\cite{Altmannshofer:2014cfa}. In Ref. ~\cite{Elahi:2015vzh}, we discussed the potential LHC reach using a dedicated cut-and-count $Z'$ analysis in the $pp \to Z \to 4\mu$ (for $M_{Z'} > 2m_{\mu}$) and $pp \to \mu^+ \mu^- \met$ (for $M_{Z'} < 2m_{\mu}$) channels. These channels were proposed for their cleanliness. Additionally, since the contribution of $Z'$ is greatest when it is produced on-shell, the mass ranges of the channels were chosen such that an on-shell $Z'$ can decay to muons ($p p \to 4 \mu$) or neutrinos ($p p \to 2 \mu \met$). Although in~\cite{Elahi:2015vzh} a large region of parameter space could be uncovered with the cut-and-count method after the full $3 \ \rm ab^{-1}$ of HL-LHC run, in this paper we will show that our sensitivity can be enhanced further if we use the matrix element technique. 
    
\section{ LHC bounds on $L_\mu-L_\tau$ model using Matrix Element Method}
\label{sec:analysis}

\subsection{ Looking for $Z'$ with mass range $ 2 m_\mu < M_{Z'} < M_{Z}$ in $ p p \rightarrow Z \rightarrow 4 \mu$ }
\label{sec:fourmu}
The rare process $ Z \to 4\mu$ occurs through $Z$ boson decay into two muons, one of which radiates a neutral boson $V$ that subsequently splits into a second pair of muons. In the SM, $V$ can be an off-shell $Z$ or photon, while in the $L_\mu -L_\tau$ model, the on-shell/off-shell $Z'$ will also contribute. 

Four muons reconstructing an on-shell $Z$ boson is a clean process that has been studied extensively. Therefore, it is an ideal channel for constraining $L_\mu -L_\tau$ model. Moreover, the related channels $pp \to Z \to 4 e$ and $pp \to Z\to 2e \ 2\mu$ can be used as a background control sample to suppress the experimental uncertainties of this channel. Within the SM the cross section of $Z \to 4\mu$ is the same as the $Z\to 4 e$ and $Z \to 2e \ 2\mu$ processes up to $O(m_e^2/m_\mu^2) \sim 10^{-4}$, hence by  measuring $Z \to 4 e $ and $Z \to 2e\,2\mu$, we can obtain a precise prediction of $(pp \to Z \to 4\mu)_{SM}$. As such, in the following, we will assume that the systematic uncertainties of $(pp \to Z \to 4\mu)_{SM}$ are sub-percent.

To study the $L_{\mu}-L_{\tau}$ $Z'$, we generated a Universal FeynRules Output (UFO) model~\cite{Degrande:2011ua} using Feynrules~\cite{Alloul:2013bka}. We then fed the model to MadGraph~\cite{Alwall:2011uj} to generate our events at leading-order. The background sample only contains the SM contributions to the four muon production $(p p \to 4 \mu)_{SM}$, while the signal event sample include both the $Z'$ as an intermediate state (where its width is calculated using MadGraph), \emph{and} the SM gauge bosons $(p p \to 4 \mu)_{SM+Z'}$ to capture the interference among processes. The signal events were generated for various values of $\alpha^{\text{gen}} = (M_{Z'}^{\text{gen}} , \epsilon^{\text{gen}} )$ between $2\,m_{\mu} \leq M_{Z'}^{\text{gen}} \leq 40\, \gev$ and $0.001 \leq \epsilon^{\text{gen}} \leq 0.1$.
 
Before applying MEM, we impose some preliminary cuts to ensure the events have been triggered upon. Specifically, we impose a di-lepton trigger used in LHC-13~\cite{Khachatryan:2016fll} that selects events with $p_T\ (\mu_1) > 17 \ \gev$ and $p_T\ (\mu_2) > 8 \ \gev$, where $\mu_1$ is the leading muon and $\mu_2$ is the sub-leading one. We also require $p_T > 4 \ \gev$ for all muons and the separation $ \Delta R_{\mu \mu} > 0.05$ ~\cite{Khachatryan:2016fll, CMS:2012bw, Aad:2014wra}. To ensure $ \sqrt{\hat s} \sim M_{Z}$, we impose $76 \ \gev < m_{4 \mu} < 106 \ \gev$ and veto extra jets in the event, since they would hurt the cleanliness of the process. 

As lepton momenta are accurately measured at the detector, we can be confident that the detected momenta very closely represents the partonic level momenta; stated in terms of transfer functions introduced in Eq.~(\ref{eq:definition}), we will assume $W(p^{\rm rec}, p^{\rm par}) = \delta (p^{\rm rec} - p^{\rm par})$. Furthermore, because the final states are fully reconstructable and we have vetoed jets\footnote{ we have done our analysis at parton level, but given $ m_{4 \mu} \sim M_{Z}$, it is reasonable to assume jet contamination is negligible.} 
, we can find the initial state energies (or equivalently the $x_i$s in Eq.~(\ref{eq:definition1})) from conservation of energy and momentum, as shown in Eq.~(\ref{eq:deltas}). As a result, all of the integrations collapse due to delta functions, and the likelihood ratio becomes 
\beq
\Lc (p^{\text{rec}};M_{Z'}, \epsilon) = \frac{ \frac{1}{\sigma} \frac{f(x_{_1}) f(x_{_2})}{2s x_{_1} x_{_2}} | \Mc(p^{\rm rec};M_{Z'}, \epsilon )_{q(x_{_1})q(x_{_2}) \to 4 \mu)}|^2}{ \frac{1}{\sigma} \frac{f(x_{_1}) f(x_{_2})}{2s x_{_1} x_{_2}}\, \, | \Mc(p^{\rm rec};SM)_{q(x_{_1})q(x_{_2}) \to 4 \mu)}|^2}.\nonumber
\eeq

Because the initial state quarks (with energies parameterized by $x_{_1}$ and $x_{_2}$) are the same for both the signal and the background at leading order, the PDF and $x_i$ factors cancel in the ratio. Hence, the likelihood ratio for a single event can be simplified to 
\beq
\Lc (p^{\rm rec};M_{Z'}, \epsilon) = \frac{ | \Mc(p^{\rm rec};M_{Z'}, \epsilon )|^2}{| \Mc(p^{\rm rec};\rm SM)|^2}
\label{eq:likelihood}
\eeq

The analytical calculation of $|\Mc|^2$ can be simplified if we just consider the process $ Z \to 4\mu$, where $Z$ is the vector sum of the four muons. Although the events could also be due to the processes $qq \to \gamma^* \to 4 \mu$ or the interference of $\gamma^*$ and $Z$ mediators, requiring $76 \ \gev < m_{4 \ell} < 106 \ \gev$ assures us $|\Mc_{Z \to 4 \mu}|^2$ is a reasonably accurate description of the events. For the analytic evaluation of $|\mathcal M|^2$, we use the form given in MCFM~\cite{Campbell:1999ah, Campbell:2011bn,Boughezal:2016wmq}, which is easily modified to include $Z'$ intermediate states (for the case of the signal).
         
Even when all final states particles are reconstructed, there are complications in evaluating $|\mathcal M|^2$. Specifically, at the detector level, we do not know the `right' pair of muons that reconstruct the $V=\gamma/Z/Z'$, leading to a combinatorics problem. The most naive way we can account for this problem is by summing over all four possible $|\Mc|^2$ with different combinations of muon pairs originating from $V$. With a better algorithm, the sensitivity to signal may be further improved, but we do not attempt that here. 

The likelihood ratio, as shown in Eq.~(\ref{eq:likelihood}) can be calculated for each event as a function of $M_{Z'},$ and $\epsilon$. Before proceeding, we need to differentiate between the analysis $\alpha = (M_{Z'}, \epsilon)$ values used in the signal matrix element hypothesis ($\mathcal M(p^{\text{rec}}, M_{Z'}, \epsilon$)) and the ``truth'' values -- the values of $M_{Z'}$ and $\epsilon$ used to generate the signal events (which, in actual data, would be unknown). To avoid confusion, we will use $\alpha^a = (M_{Z'}^a,\epsilon^a$) for the analysis values and $\alpha^{\text{gen}} = (M^{\text{gen}}_{Z'},\epsilon^{\text{gen}})$ which were used for event generation. 

Our goal is to determine the values of $M_{Z'}^a$ and $\epsilon^a$ that maximize our sensitivity to the signal hypothesis.
Keeping this task in mind, it is worth looking at how the $\alpha^a$ parameters enter into the matrix element: 

\beq 
 \Mc(p^{\text{rec}};M_{Z'}^a, \epsilon^a )= \Mc_{\rm SM} + \frac{( \epsilon^a)^2}{M_{\mu \mu}^2 -( M_{Z'}^a)^2 - i \Gamma_{Z'}^a M_{Z'}^a}(\cdots),
\label{eq:NP}
\eeq
where $\Gamma_{Z'}^a= \Gamma_{Z'}(M_{Z'}^a, \epsilon^a)$ and the ellipses represent the part of the matrix element that is independent of $M_{Z'}^a, \epsilon^a$. 
Knowing the dependence of $\Mc$ on $\alpha^a$ can help us better understand the behavior of the likelihood ratio for various $\alpha^a$ values. For example, the appearance of $M_{Z'}^a$ only in the denominator suggests that the likelihood ratio is very sensitive to the value of $M_{Z'}^a$ (e.g, the possibility of resonance). With $\epsilon^a$ in the numerator, we suspect the likelihood ratio to not be as sensitive to a particular value of $\epsilon^a$, because regardless of which testing hypotheses the events belongs to, their likelihood ratio will increase by increasing $ \epsilon^a$. Although $\Gamma_{Z'}^a$ in the denominator also depends on $\epsilon^a$, its contribution is suspected to be much smaller than the $\epsilon^a$ in the numerator. 

To examine these conjectures and hunt for the $\alpha^a$ that maximize the likelihood ratio, we will proceed by plotting $\sum_i \log[\Lc(p_i, \alpha^a)]$ defined in Eq.~(\ref{eq:plot}) as a function of $\alpha^a$. Thereby, for each $\alpha^a$ we can compare the relative sizes of the summed log likelihood ratio of the signal sample events with that of background sample, and look for the values of $\alpha^a$ that maximize the likelihood of the signal with respect to the background. 
As we have two new parameters ($\alpha^a = M_{Z'}^a, \epsilon^a$), for simplicity we will fix one of the parameters and plot Eq.~(\ref{eq:plot})) as a function of only one parameter.

\begin{figure}[h!]
\centering 
\includegraphics[width=0.49\textwidth, height=0.27\textheight]{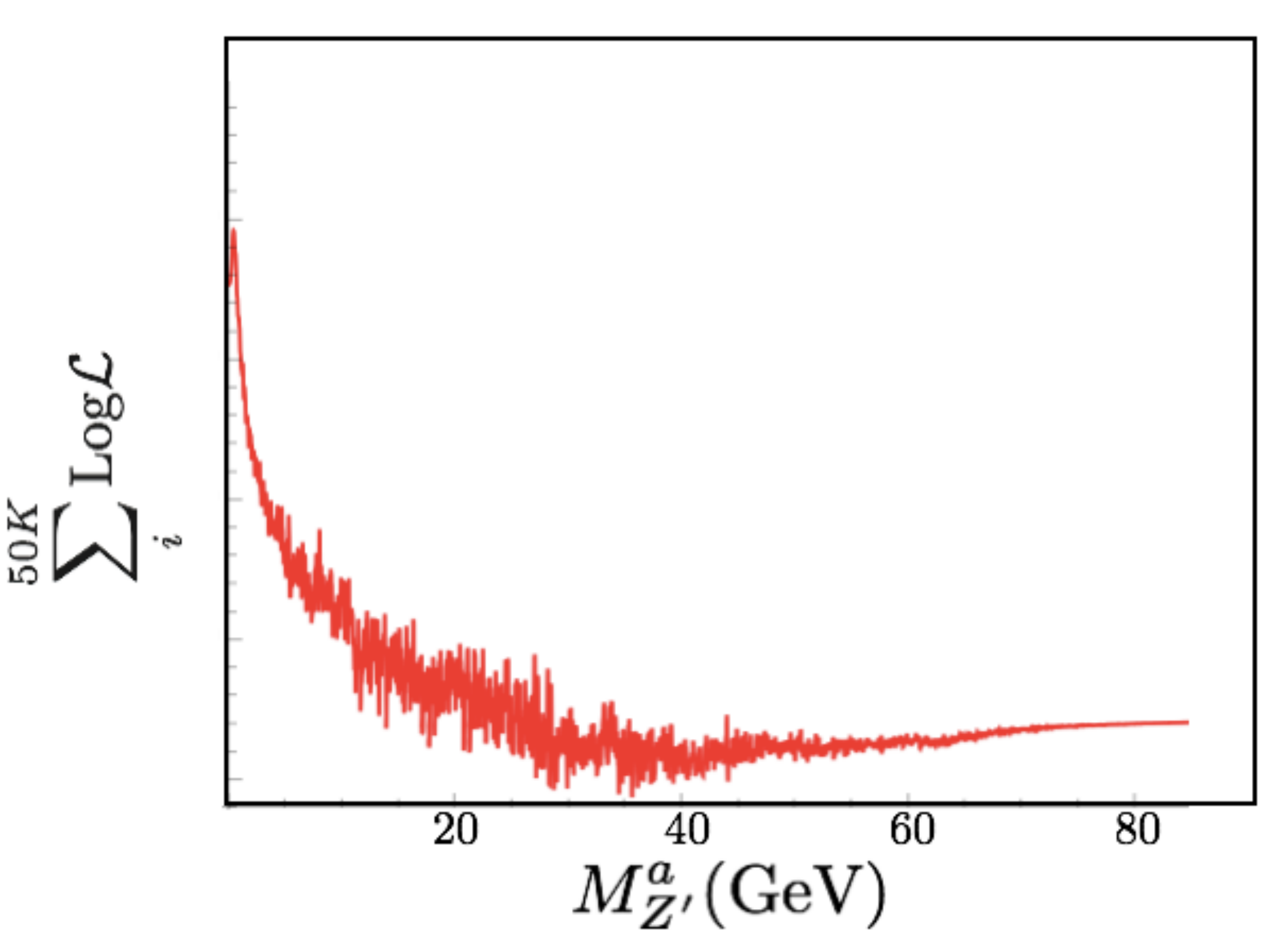}
\includegraphics[width=0.5\textwidth, height=0.27\textheight]{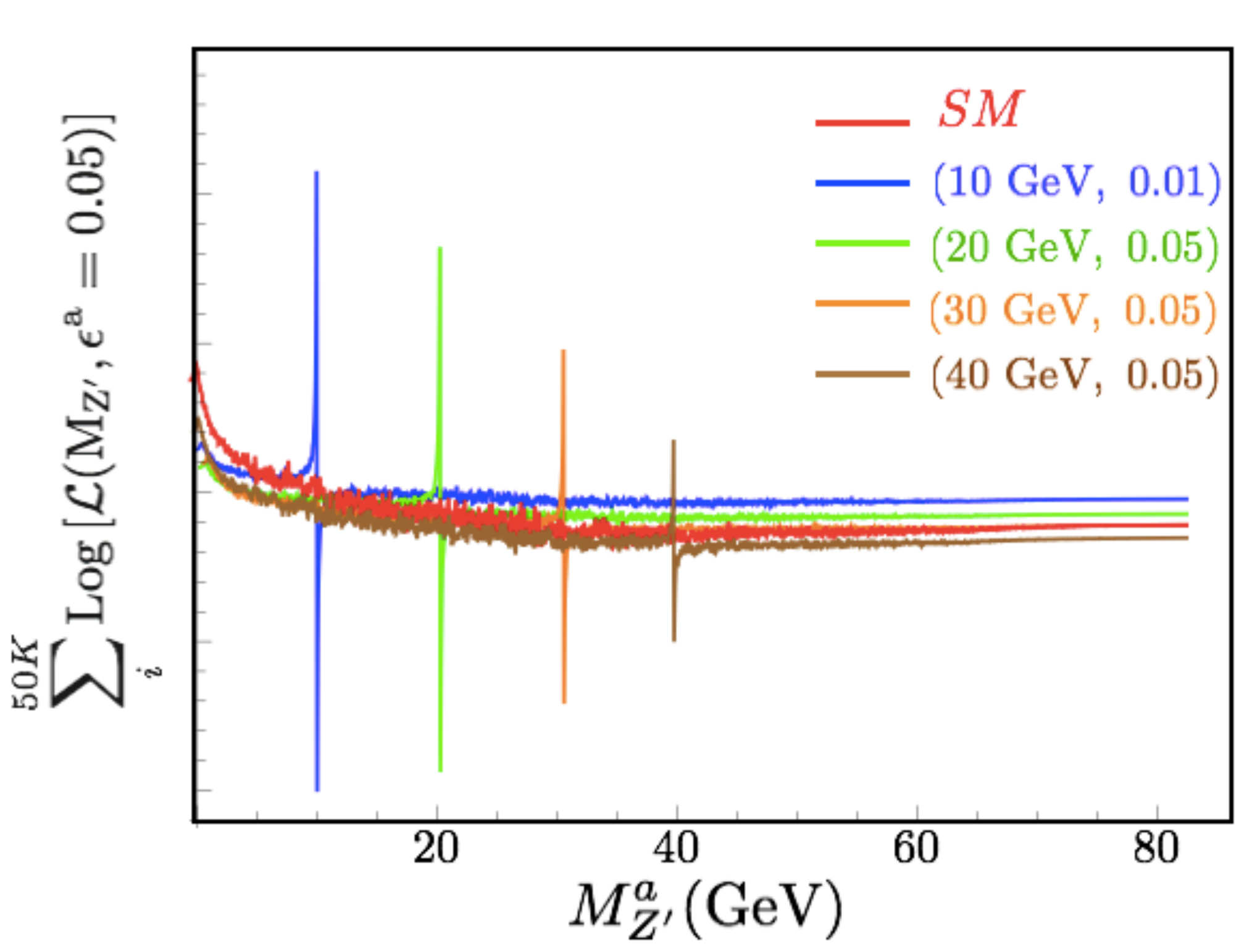}
\caption{The plot of log likelihood ratio, summed over 50000 events is shown above. The left plot is the distribution of SM events, where there is a spike at every reconstructed $M_{\mu \mu}$. The peak at $M_{Z'}^a \sim 0 \ \gev$ is due to the photon propagator. For larger values of $M_{Z'}^a$, there are fewer events with $M_{\mu \mu} = M_{Z'}^a$, and the function looks smoother, which is the result of our basic cuts. In the right plot, the distribution of the signal events compared to the simulated background events are shown. In the signal events, which include both SM and $Z'$ contributions, there is a peak at $M_{Z'}^a = M_{Z'}^{\rm gen}$ because of the contribution from the on-shell $Z'$. It is important to mention the optimal value of $M_{Z'}^a$ is obtained by comparing the relative shapes of the signal plot with the background one. The scale on the y-axis is irrelevant because it is highly sensitive to the number of events in our sample, and by increasing the sample size the numerical values on the y-axis of the plots becomes more comparable. For our analysis, we manually brought the plots to similar y-axis values, to compare their shapes.}
\label{fig:maxlikelihood}
\end{figure}

First, we assume a non-zero value for $\epsilon^a $ and plot Eq.~(\ref{eq:plot}) for a range of $M_{Z'}^a$ values. The value of $\epsilon^a$ can be any arbitrary non-zero value, and varying $\epsilon^a$ does not alter our results. For every event, the reconstructed invariant mass of two muons ($M_{\mu \mu}$) will produce a spike at $M_{Z'}^a= M_{\mu \mu}$, which is not necessarily equal to the $M_{Z'}^{\text{gen}}$. In fact, because we are summing over all four possible combinations of muon pairs, we have four spikes in $M_{Z'}^a$ for each event. Therefore, summing over all events,  the net distribution (in our case, 50000 events) has spikes at all values of $M_{Z'}^a \lesssim M_{Z}$ as shown in the left plot of Fig.~\ref{fig:maxlikelihood}. In the signal sample, due to the higher number of events with $M_{\mu \mu}= M_{Z'}^{\rm gen}$, the spike at $M_{Z'}^{\rm gen}$ is more noticeable (see Fig.~\ref{fig:maxlikelihood}, right plots). The maximum likelihood ratio of the signal events is consistently at $M_{Z'}^a =M_{Z'}^{\rm gen}$ for any arbitrary value of $\epsilon^a \neq 0$, while the pure background events show no interesting behavior at $M_{Z'}^a =M_{Z'}^{\rm gen}$.

The height of the peak at $M_{Z'}^a = M_{Z'}^{\rm gen}$ relative to the other spikes depends on $\epsilon^{\rm gen}$ (regardless of $\epsilon^a$). The peak is more visible for larger values of $ \epsilon^{\rm gen}$, but becomes less distinguishable from other spikes for smaller values of $\epsilon^{\rm gen}$. As one example, for $ \epsilon^{\rm gen} \lesssim 0.01$ for $M_{Z'}^{\rm gen} \gtrsim 10\, \gev$, we could not distinguish the peak at $M_{Z'}^a = M_{Z'}^{\rm gen}$ from other spikes).
Given that our analysis highly depends on whether we are able to find $M_{Z'}^{\rm gen}$ by plotting $\sum_i \log[\Lc(p_i, M_{Z'}^a, \epsilon^a_{\text{ arbitrary}} \neq 0 )]$ as a function of $M_{Z'}^a$, our strategy needs to bifurcate depending on the value of $ \epsilon^{\rm gen}$:
\begin{itemize}
\item large $\epsilon^{\text{gen}}$: we determine the value of $M_{Z'}^a$ by observing the peak at $M_{Z'}^a = M_{Z'}^{\rm gen}$;
\item small $\epsilon^{\text{gen}}$: we need to chose an arbitrary value of $M_{Z'}^a$. 
\end{itemize}
However, as the choice of the strategies depends on whether we are able to \textit{observe} the peak at $M_{Z'}^a = M_{Z'}^{\rm gen}$, rather than trying to quantify what it takes to observe a peak, we will study both strategies for all values of $\epsilon^{\rm gen}$. More specifically, we will first assume that we can find $M_{Z'}^a =M_{Z'}^{\rm gen}$ even for small values of $ \epsilon^{\rm gen}$, then we will fix an arbitrary value for $M_{Z'}^a$ and study the signal assuming we cannot find $M_{Z'}^{\rm gen}$ even for large values of $ \epsilon^{\rm gen}$. The first method represents the best we can do with the MEM, while the second represents a more realistic reach for $\epsilon \lesssim 0.01$.

\begin{figure}[h!]
\centering 
\includegraphics[width=0.5\textwidth, height=0.27\textheight]{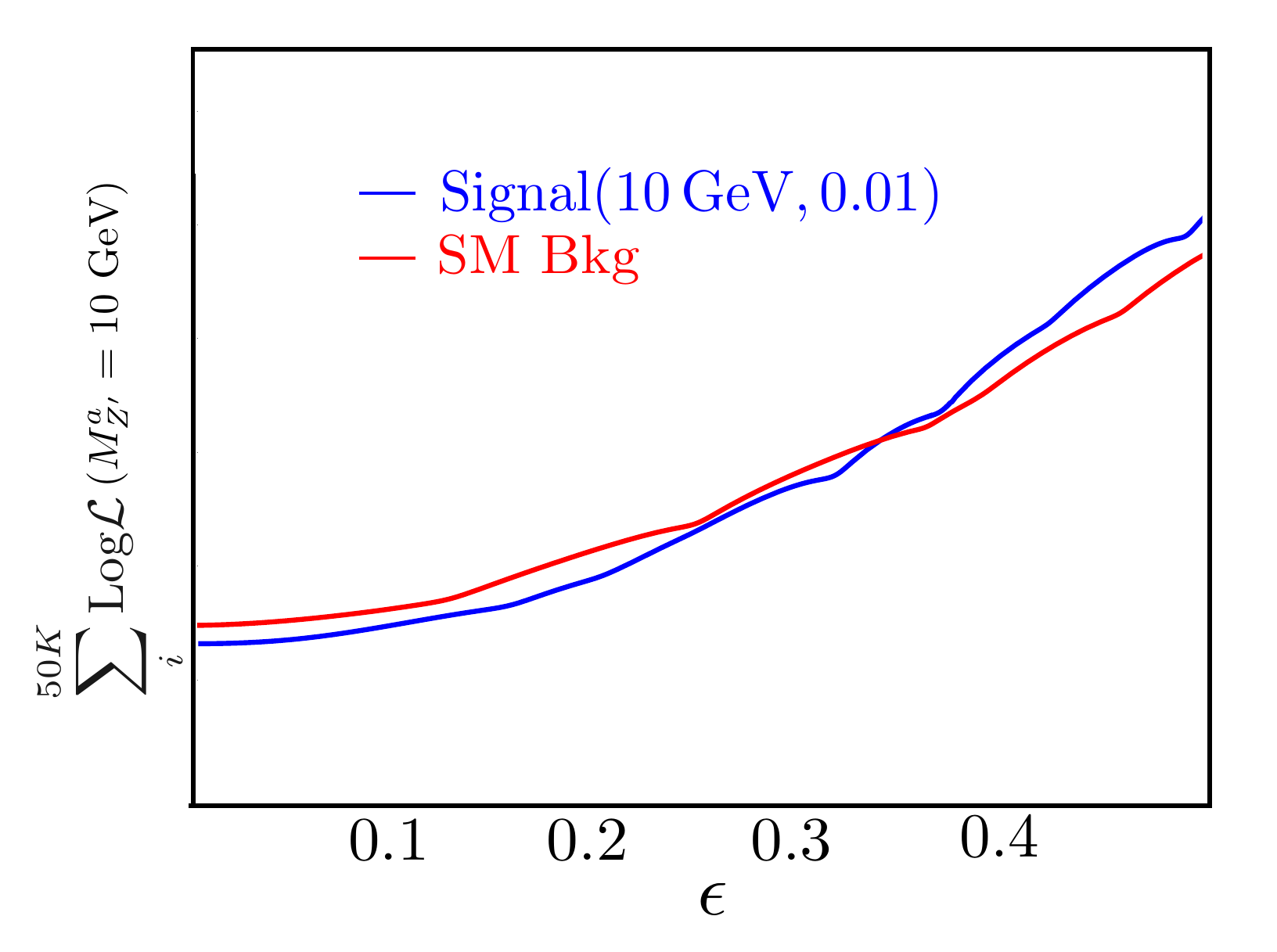}
\caption{ The distribution of log likelihood ratio for the signal and background events summed over 50000 sample events. We have fixed $M_{Z'}^a = M_{Z'}^{\rm gen} = 10\, \gev$ while $\epsilon^{\rm gen} = 0.01$. The signal and the background MC sample events increase as $\epsilon^a$ increases, providing us with no special $ \epsilon^a$ value that optimizes the signal likelihood ratio. The numerical value on the y-axis of the two plots as explained earlier is irrelevant and thus not shown. 
}
\label{epsilon}
\end{figure}

Now that we have discussed how to find the optimal value of $M_{Z'}^a$, we follow the same procedure to determine $\epsilon^a$. For simplicity, we fix $M_{Z'}^a = M_{Z'}^{\rm gen}$ and let $\epsilon^a$ be a free parameter. The $\sum_i \log[\Lc(p_i, M_{Z'}^a=M_{Z'}^{\rm gen}, \epsilon^a )]$, where the sum is over $50000$ MC generated events for each sample, are plotted in Fig.~\ref{epsilon} below as a function of $\epsilon^a$. Both the signal and background MC events increase as a function of $\epsilon^a$. Such behavior is anticipated, because as we can see in Eq.~(\ref{eq:NP}), larger $\epsilon$ results in larger $|\Mc(p^{\rm rec};M_{Z'}, \epsilon )|^2$ and thus larger likelihood ratio for any event. Stated another way, fixing $M_{Z'}^a$ and studying the likelihood ratio with respect to $\epsilon^a$ is not useful because the simulated signal and background events behave the same way as a function of $\epsilon^a$. Thus, for the remainder of this section, we will explore an alternative different technique for estimating the optimal $\alpha^a$.

The problem with the approach of $\Lc$ maximization is that, for a given $\alpha^a$, it compares only two numbers: the summed log likelihood ratio of the simulated signal events vs. that of background events. With this approach, we are throwing away a lot of information about each event. We can get a better handle on the signal vs. the background, if we look at their \emph{distributions} for a given $\alpha^a$ and how those distributions evolve as we vary $\alpha^a$. The distribution will illustrate to us the behavior of events for a given $\alpha^a$, rather than just their sum. We can use this distribution to find $\alpha^a$ that results in best discrimination of the signal from the background, even if the sum over events is similar. The signal usually resides in high values of the likelihood ratio, as expected according to the definition of the likelihood ratio (Eq.~(\ref{eq:likelihoodratio})). So, we distribute the events in the inverse likelihood ratio ($\Lc^{-1}$) to be able to see the excess of the signal spread over a narrower window $[0,1]$.\footnote{not all of the background will fall in this $[0,1]$ region, however we only care about how the background is distributed in the region where the signal resides.} 
 
\begin{figure}[h!]
\centering 
\includegraphics[width=0.65\textwidth, height=0.4\textheight]{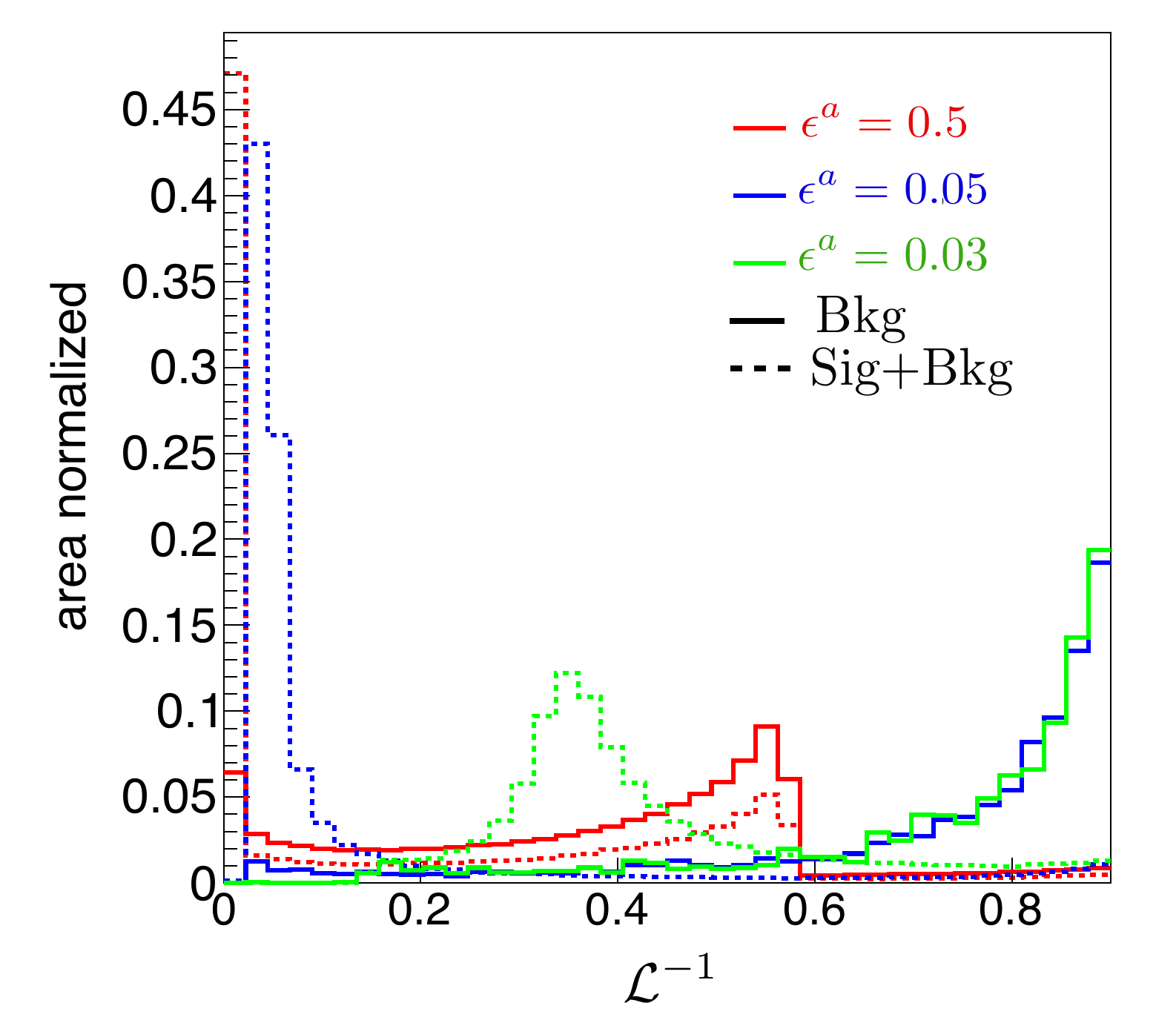}
\caption{The distribution of $\Lc^{-1} = |\Mc_{\rm SM}|^2/| \Mc_{Z'+\rm SM}|^2 $ for $M_{Z'}^{\text{gen}} = 10 \ \gev,$ and $\epsilon^{\text{gen}} = 0.05$. This plot shows that for $M_{Z'}^a = 10 \ \gev$, if we have $\epsilon^a = 0.03$, the resonance region is not very well-separated from background region. If $ \epsilon^a = 0.5$, although the signal is far from 1, but the background also spreads more. The middle value $ \epsilon^a = 0.05$ is the optimal one. 
}
\label{fig:epsa1}
\end{figure}

As an example of $\alpha^a$ distribution method, in Fig.~\ref{fig:epsa1} below we show the distribution of the MC generated events for the signal and the background for different choices of $\epsilon^a$, assuming $M_{Z'}^a = M_{Z'}^{\rm gen}$. We can see that as we increase $\epsilon^a$, the resonance region -- defined as events with $M_{\mu \mu} \sim M_{Z'}^{\rm gen}$ -- is more separated from the background. However for very large $\epsilon^a$ (for example $\epsilon^a = 0.5$ in the Fig.~\ref{fig:epsa1}), the background distribution also becomes more spread. This is again consistent with Eq.~(\ref{eq:NP}), because no matter whether the events are at resonance or not, increasing $\epsilon^a$ increases their likelihood ratio. 

The optimal value of $ \epsilon^a$ depends on the distribution of the signal with respect to the background. For a fixed sample size of signal and background simulated events, optimal is defined as maximization of 
\beq 
\frac{\text{number of signal events}}{\sqrt{\text{number of background events}}}.\nonumber 
\eeq 
After examining a few values\footnote{The benchmark points $\epsilon^a$ we studied are $\epsilon^a \in [0.01, 0.5]$ with increments of $0.01$. With a larger sample of $\epsilon^a$, the best value of $\epsilon^a$ may slightly vary.} of $\epsilon^a$, we find $\epsilon^a = 0.05$ to give the best discrimination, regardless of the $\epsilon^{\rm gen}$. This result is consistent for all values of $M_{Z'}^{\rm gen}$ that we studied. 
                
\begin{figure}[h!]
\centering 
\includegraphics[width=0.65\textwidth, height=0.37\textheight]{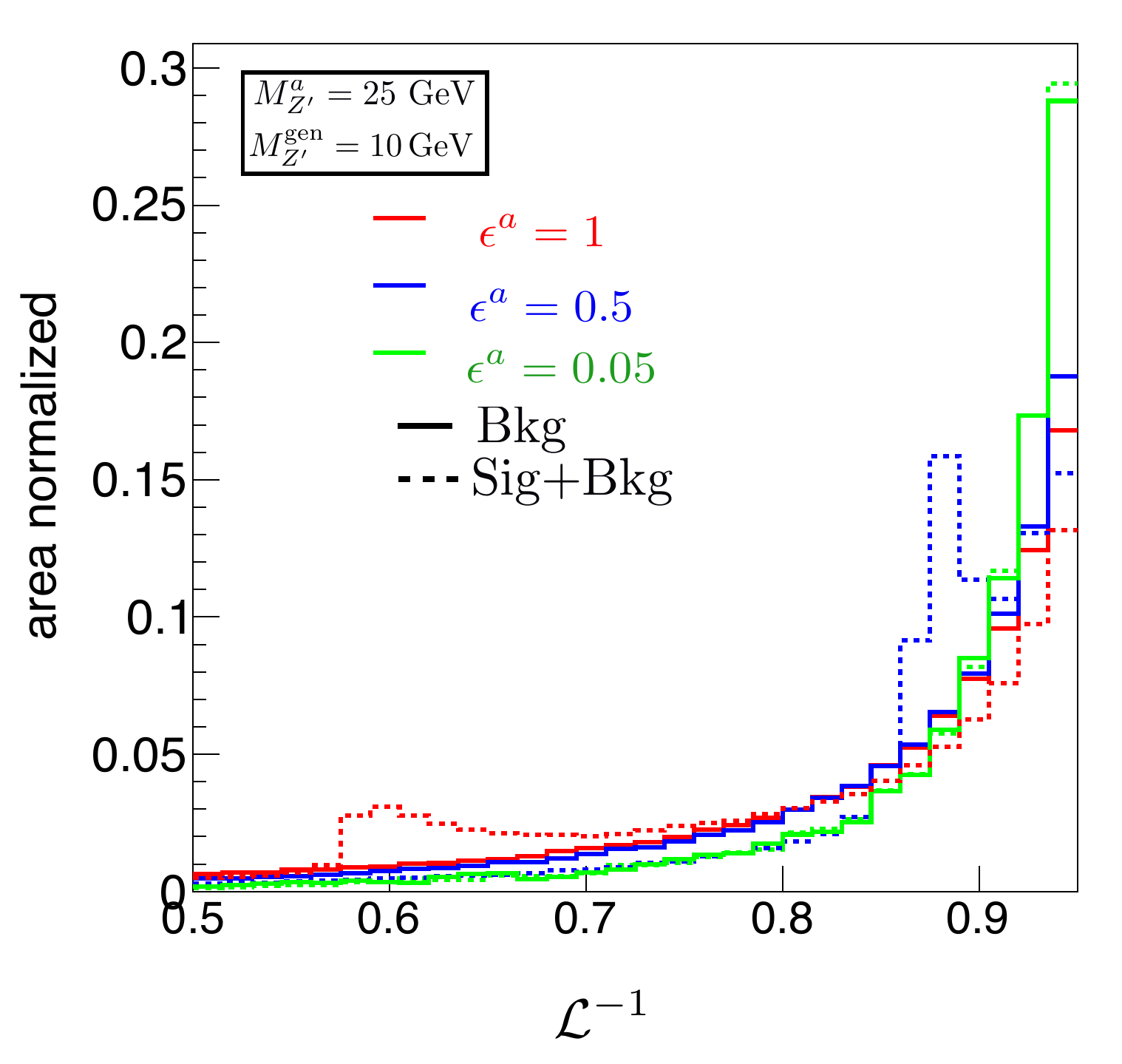}
\caption{The distribution of $\Lc^{-1} = |\Mc_{\rm SM}|^2/| \Mc_{Z'+\rm SM}|^2 $ for $M_{Z'}^{\rm gen} = 10 \ \gev,$ and and $\epsilon^{\rm gen} = 0.05$ is shown, where we are assuming $M_{Z}^a = 25 \ \gev$, and we vary $ \epsilon^a$. Here, a relatively large $ \epsilon^a$ is needed to separate the signal from background. However, with increasing $\epsilon ^a$, we also get broader peaks and therefore our sensitivity is not good as when $M_{Z'}^a = M_{Z'}^{\rm gen}$. In this particular example $ \epsilon^a = 0.5$ is the optimal choice.}
\label{fig:epsa2}
\end{figure}

Thus far, we have seen that with the assumption that we know $M_{Z'}^{\rm gen}$, the optimal values for $\alpha^a$ are $M_{Z'}^a = M_{Z'}^{\text{gen}}$ and $\epsilon^a = 0.05$. 
However, if we cannot find the value of $M_{Z'}^{\rm gen}$ by maximizing $\sum_i \log[\Lc(p_i, \alpha^a)]$ with respect to $M_{Z'}^a$, we must explore how our sensitivity changes if $M_{Z'}^a$ is fixed to an arbitrary value $\ne M_{Z'}^{\rm gen}$. 
Because the $Z'$ mass for the generated events is no longer the same as the $Z'$ mass used in the analysis matrix element, the events with $M_{\mu \mu} \sim M_{Z'}^{\rm gen}$ are no longer at resonance in the matrix element $\Mc(p^{\rm rec};M_{Z'}, \epsilon )$ (Eq.~(\ref{eq:NP})). Therefore,  when we go to search for the optimal $\epsilon^a$ using distributions, the signal, and background are less separated than in the previous case. To increase the separation, we need to increase $\epsilon^a$.
Consequently, the optimal $ \epsilon^a$ is no longer fixed at $0.05$ and will depend on the difference between $M_{Z'}^a$ and $M_{Z'}^{\rm gen}$. 
For example, in Fig.~\ref{fig:epsa2}, the distribution of the signal and the background MC generated events for $M_{Z'}^a = 25 \ \gev$ and $M_{Z'}^{\rm gen} = 10 \ \gev$, with $\epsilon^a = 0.05,\, 0.5,\,1$ are shown. 
In this example, we see $\epsilon^a = 0.05$ does not give a good discrimination of the signal from the background, and we have to use larger values of $\epsilon^a$.
Furthermore, as the plot illustrates, although large $\epsilon^a$ increases the separation between the signal region and the background, it will broaden the signal region.
Because we are interested in distinguishing the region where the signal to background ratio is maximized, having a broad signal region is not ideal. 
Hence, we expect our reach for $M_{Z'}^a \neq M_{Z'}^{\rm gen}$ to be more limited that the previous case, even with the most optimal $\epsilon^a$. 

Table~\ref{tab:summary} recapitulates the value $\epsilon^a$ depending on whether or not plotting $\sum_i \log[\Lc(p_i, \alpha^a)]$ as a function of $M_{Z'}^a$ can tell us $M_{Z}^{\rm gen}$. 
\begin{table}[h!]
\begin{center}
\begin{tabular}{l| r c }
\hline
\hline

$M_{Z'}^a = M_{Z'}^{\rm gen}$& $\epsilon^a = 0.05$\\
 using likelihood maximization &\\
 \hline
$M_{Z'}^a$ arbitrary & $ \epsilon^a > 0.05$ \\
 likelihood maximization fails, because of small $\epsilon^{\rm gen}$ & \\
\hline
\hline
\end{tabular}
\caption{Determining $\alpha^a = (M_{Z'}^a, \epsilon^a)$ using a combination of likelihood maximization and the $\Lc^{-1}$ distribution.}
\label{tab:summary}
\end{center}
\end{table}

\begin{figure}[h!]
\centering 
\includegraphics[width=0.7\textwidth, height=0.37\textheight]{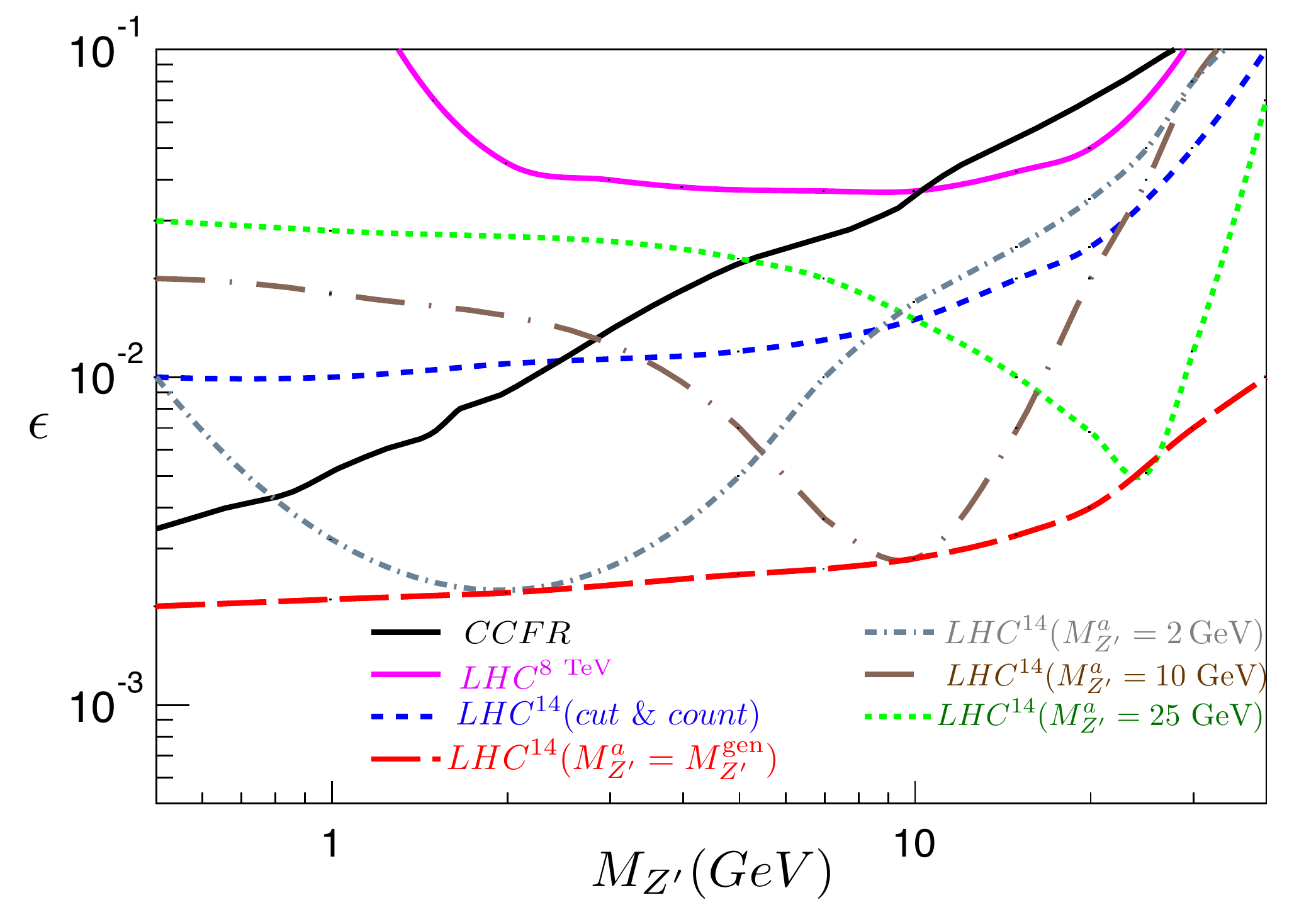}
\caption{The bound on $L_\mu-L_\tau$ from different experiments. The purple line is the LHC-8 bounds, and the black line is CCFR bound. The dashed blue line is the bound from LHC-14 with luminosity $ 300 \ \rm fb^{-1}$ up to $ 3\sigma$ with cut and count method~\cite{Elahi:2015vzh}. The red dashed line in the figure shows the reach with MEM, but assuming we know the mass of $Z'$, and with $\epsilon^a = 0.05$. This is the best sensitivity we could get using MEM. The dotted green line is assuming $M_{Z'}^a = 25\ \gev$, the dashed brown line is $M_{Z'}^a = 10\ \gev$, the dotted-dashed gray line is $M_{Z'}^a = 2\ \gev$ while choosing the optimal value for $\epsilon^a$. The MEM bounds are also with luminosity $300 \ \rm fb^{-1}$ up to $ 3\sigma$. }
\label{fig:epsmzp}
\end{figure}

To quantify the reach of the strategy proposed here, let us define the following variables: 
\begin{align*}
S &= \text{Luminosity} \times ( \sigma(pp \to 4 \mu)_{\rm SM+Z'} - \sigma(pp \to 4 \mu)_{\rm SM})\\
B &= \text{Luminosity} \times ( \sigma(pp \to 4 \mu)_{\rm SM}),
\end{align*}
The cuts on $\Lc^{-1}$ are imposed such that we get the maximum $S/\sqrt{B}$ with integrated luminosity of $300 \ \rm fb^{-1}$. We also require $S> 10$ to avoid confusion of signal events for statistical fluctuations. We find that, by using the MEM approach we are able to get $S/\sqrt{B} \geq 3$ up to $\epsilon^{\rm gen} \sim 0.002$ and $M_{Z'} \lesssim 20 \ \gev$, and up to $ \epsilon^{\rm gen} \sim 0.005$ for $ 20 \, \gev < M_{Z'}^{\rm gen} \lesssim 40 \, \gev$, provided that we can determine the true value of $M_{Z'}^{\rm gen}$ by conventional MEM means. This best case scenario is indicated with the dashed red line in Fig.~\ref{fig:epsmzp}.
The dashed-dotted gray line in Fig.~\ref{fig:epsmzp} shows our reach assuming $M_{Z'}^a = 2\ \gev$ while optimizing $\epsilon^a$. Similarly, the dashed brown line is for $M_{Z'}^a = 10\ \gev$, and the dotted green line is for $M_{Z'}^a = 25\ \gev$. As expected, the lines with arbitrary $M_{Z'}^a$ touch the dashed red line for $M_{Z'}^a = M_{Z'}^{\rm gen}$, and have a relatively good sensitivity when $M_{Z'}^a\, \sim\, M_{Z'}^{\rm gen}$, but their sensitivity declines as $M_{Z'}^a$ moves away from $M_{Z'}^{\rm gen}$. The current constraints from CCFR experiment and LHC-8 are also shown in Fig.~\ref{fig:epsmzp} in black and purple solid lines, respectively. The expected exclusion bound (3 $\sigma$) from LHC-14, with $300 \ \rm fb^{-1}$ integrated luminosity, using the cut-and-count method that is explored in Ref.~\cite{Elahi:2015vzh} is the dashed blue line in Fig.~\ref{fig:epsmzp}. With the assumption that we know $M_{Z'}^a = M_{Z'}^{\rm gen}$, our reach with MEM is about a factor of 10 greater than the canonical cut-and-count method, and even with $M_{Z'}^a$ fixed to an arbitrary value, we have an improved sensitivity for some range of $M_{Z'}^{\rm gen}$ compared with the cut-and-count method.  By trying several values of $M_{Z'}^a$ and choosing $ \epsilon^a \sim 0.05$, one can achieve a sensitivity near the red curve shown in Fig.~\ref{fig:epsmzp} at the LHC. 
            
\subsection{ Looking for $Z'$ with mass range $ M_{Z'} < 2 m_\mu $ in $ p p \rightarrow \mu^+ \mu^-$ \MET}
\label{sec:twomu}
    If $M_{Z'} < 2 m_\mu$, the decay of on-shell $Z'$ to muons is kinematically forbidden. In fact, for this mass range, on-shell $L_\mu-L_\tau \ Z'$ can only go to neutrinos. While we could hunt for off-shell $Z'$ in this mass range using the $4\mu$ final state (as in the previous section), off-shell $Z'$ production is suppressed relative to on-shell production by two additional powers of $\epsilon$. Therefore, as proposed in~\cite{Elahi:2015vzh}, we will give up the benefits of a completely visible final state in favor of larger rate and hunt for $Z'$ in $p p\to \mu^+ \mu^- \met$. The signal contribution to $pp \to \mu^+\mu^- + \met$ comes from the production of a pair of muons, one of which radiates an on-shell $Z'$ that decays to a pair of neutrinos~\footnote{The signal also captures the diagram where the neutrinos and muons are produced in opposite order: a pair of neutrinos are first produced, and then one of the neutrinos emits a $Z'$ which splits to two muons. In this topology, the $Z'$ must be off-shell, so the  cross section is suppressed by more powers of $\epsilon$ and is, therefore, negligible in the parameter space of interest.}: $p p \to \mu^+\mu^- Z' \to \mu^+ \mu^- \nu_\ell \bar \nu_\ell$ . 
    
    Due to the presence of missing energy, we can no longer impose an invariant mass cut on the final state. As a consequence, some new, important backgrounds emerge:
\begin{align}
pp\rightarrow&\, \left.\tau^+ \tau^- \right|_{\text{dimuon decay}}, \nonumber \\
pp\rightarrow&\, V V = \left\{ \begin{matrix}\left. W^+ W^- \right|_{\text{dimuon decay}},\\ \left. Z \,(Z/\gamma)\right., \ \ \ \ \ \ \ \ \ \ \ \ \ \ \ \ \ \end{matrix} \right. \nonumber\\
pp \rightarrow&\, \mu^+\mu^-+ \text{jets}.
\label{eqn:backgrounds}
\end{align}

The last background arises as a result jet mis-measurements and pileup, and so peaks at low values of \MET. Di-tau production is the largest irreducible background, followed by diboson production. Technically, we include both resonant and non-resonant contributions in this category, as the latter can be non-negligible. Thus, the $VV$ background is more accurately described as $(pp \to \mu^+\mu^-\nu_{\ell}\bar{\nu}_{\ell})_{SM}$. Similarly, the `signal' in this section is defined as $pp \to \mu^+\mu^-\nu_{\ell}\bar{\nu}_{\ell}$ including $Z'$ as a possible intermediate state: $(pp \to \mu^+\mu^-\nu_{\ell}\bar{\nu}_{\ell})_{SM+Z'}$. As in the previous section, the signal is defined including SM contributions to incorporate interference. 

To study this channel in more detail, we generated events for $pp \to \tau^+ \tau^- \to \mu^+\mu^- + \met$, $(pp \to \mu^+\mu^-\nu_{\ell}\bar{\nu}_{\ell})_{SM}$, and $(pp \to \mu^+\mu^-\nu_{\ell}\bar{\nu}_{\ell})_{SM+Z'}$ via the MC chain {\tt MadGraph5-aMC@NLO}~\cite{Alwall:2011uj} plus {\tt Pythia 6.4} ~\cite{Sjostrand:2006za}, where the latter step is used here to decay the taus.\footnote{We restricted the possible $\tau$ decays to leptonic channels only within {\tt Pythia 6.4} to make event generation more efficient} Before any MEM analyses, we require all events to pass the same dilepton trigger requirement used in Sec. \ref{sec:fourmu} ($p_T\ (\mu_1) > 17 \ \gev$ and $p_T\ (\mu_2) > 8 \ \gev$, where $\mu_1$ is the leading muon and $\mu_2$), veto any jets with $p_T > 20 \, \gev, \eta <2.5$, and impose a minimum missing energy cut of $\met > 20\, \gev$. The last cut is imposed to suppress the $pp \to \mu^+\mu^- + \text{jets}$ background.

    After the initial set of cuts, muonic tau production $ p p \to \tau^+ \tau^- \to \mu^+ \mu^- \met$ is our main irreducible background. The cross section of the $ \left.\tau^+ \tau^- \right|_{\text{dimuon}}$ background is roughly two orders of magnitude larger than $\sigma(pp \to \mu^+\mu^-\nu_{\ell}\bar{\nu}_{\ell})_{SM}$. Therefore, if we want to have any chance to be sensitive to a $Z'$ signal, we need to first make the $ \left.\tau^+ \tau^- \right|_{\text{dimuon}}$ background more manageable. So, instead of trying to discriminate $Z'$ signal against background, we will focus on distinguishing $ \left.\tau^+ \tau^- \right|_{\text{dimuon}}$ from other processes. 
 
 The most efficient way to eliminate the $ \left.\tau^+ \tau^- \right|_{\text{dimuon}}$ background is by using a variable that is most faithful to $ \left.\tau^+ \tau^- \right|_{\text{dimuon}}$ background and thus localizes its simulated events to a small region. Motivated by the benefits of the MEM discussed previously, we will use the $|\Mc_{\tau \tau}|^2 $, the squared matrix element of the muonic di-tau production, as the discriminating variable. For each event we will calculate $|\Mc_{\tau \tau}|^2$ using the observed final state momenta, then search for and select out regions (using MC) of $|\Mc_{\tau \tau}|^2$ that $pp \to \tau^+\tau^-$ does not populate. In doing this, we are not following the traditional MEM in this section, because we are not using $|\Mc|^2$ of the signal to distinguish that from other processes. Rather, we are only using the $|\Mc|^2$ of (part of) the background, which makes our approach independent of the signal ($Z'$) model.
 
 As a further deviation from the traditional MEM, we will weight each event by only one squared matrix element, $|\Mc_{\tau \tau}|^2 $, rather than two (a `signal' hypothesis and a `background' hypothesis). We may loose some discriminating power by not calculating the likelihood ratio as described in section \ref{sec:MEM}, but our approach is more time efficient as we do not have to deal with other squared matrix elements that contain missing energy. 
 
 Even after reducing our discriminant to the evaluation of a single $|\Mc|^2$ for each event, evaluating $|\Mc_{\tau \tau}|^2$ is still a difficult task. In $ \left.\tau^+ \tau^- \right|_{\text{dimuon}}$ production, there are four sources of missing energy, which translates to 12 unknown momenta. Moreover, because we do not know the energies of the initial state, we have in total 14 unknowns. Energy-momentum conservation $ \delta^4 (p_{\text{initial}}- p_{\text{final}})$ reduces the number of unknowns to 10, but 10 integrations for each event is still extremely computationally cumbersome. Thankfully, the specific topology of the $ \left.\tau^+ \tau^- \right|_{\text{dimuon}}$ production can help us approximate the unknowns. The list of our assumptions to approximate the unknowns are the followings:

 \begin{enumerate}
 
 \item We will assume that the $\tau$s were produced on-shell. We know that the invariant mass of an on-shell tau decay products is the tau mass. Therefore, we can determine two unknowns from this assumptions since there are two $\tau$s in each process. 
 \item Instead of calculating $\tau \to \nu_\tau \bar{\nu}_\mu \mu$, we replace the two neutrinos with one massive scalar\footnote{Technically we should also consider vector massive neutrinos, though we do not expect this choice affects our results. For $m_{\nu_s} = 0$ the vector results are identical to the scalar case, while there is a small shift in the matrix element if $m_{\nu_s} \ne 0$.} neutrino ($\nu$s for the notation), and we calculate $\tau \to \mu \ \nu$s. 

 Three body decay has different kinematic distributions compared to two body decays, but in this approach we can reduce our number of unknowns by 4, leaving 4 remaining. Because we are no longer dealing with the actual $|\Mc|^2$ and we are calculating $ p p \to \tau^+ \tau^- \to \mu^+ \mu^- \nu s \ \bar {\nu s}$, we refer to the matrix element we calculate as the ``modified'' $|\Mc_{\tau \tau}|^2$, or $|\Mc^{mod}_{\tau\tau}|^2$. 
 
 \item The tau pair can be produced from either a photon or a $Z$ boson. However, as a result of our basic cuts (di-lepton trigger and $\met > 20 \ \gev$), we can be confident that the production of $\tau^+\tau^-$ is dominated by $Z$ exchange. Therefore, we will assume that the taus are produced from an on-shell $Z$, and so $\sqrt{\hat s} = M_Z$. This assumption eliminates another unknown. 
 
 \item Based on the previous assumptions, we expect the tau decay products to be nearly collinear. Consequently, the $\eta$ and $\phi$ of the $\nu s_i$ should be close to $\eta$ and $\phi$ of the corresponding $\mu_i$, with the subscript $i$ defined as the following: $ \tau_i \to \mu_i \nu s_i$. For our analysis, we assume $ \eta (\mu_i) = \eta(\nu s_i)$, and $ \Delta \phi( \mu_i , \nu s_i) = \epsilon_\phi^i $ where $\epsilon_\phi \ll 1$. Therefore, we only work to first order in $\epsilon_\phi$. These assumptions specify two more unknowns and results in the relationship $p_T^{\tau_i} = p_T^{ \nu s_i}+ p_T^{\mu_i}.$ 
 
 \item We can also assume $p_T^{\tau_1}= p_T^{\tau_2}$. This is the same as assuming there is no initial or final state radiation, which is reasonable given that $ \sqrt{\hat s} \lesssim M_{Z}$, and we have vetoed jets in our events. This assumption leads to specification of one more unknown. 

 \end{enumerate} 
 
 Making the above approximations \footnote{ Although we have defined ``modified" specifically for the second approximation, we generalize it definition to encompass \emph{all} of the aforementioned approximations.}  , we can determine all of the unknown kinematic parameters and therefore calculate $|\Mc^{mod}_{\tau \tau}|^2$ with no integrations, {\em significantly} reducing the time and computational power needed to do the analysis. The analytical expression of $|\Mc^{mod}_{\tau \tau}|^2$ is given in Appendix~\ref{app:tataME}.
 
 We emphasize that these assumptions are only reasonable in the context of the $ \left. \tau^+ \tau^-\right|_{\text{dimuon}}$ production. For the rest of the processes ($(pp \to \mu^+\mu^-\nu_{\ell}\bar{\nu}_{\ell})_{SM}$ background and $(pp \to \mu^+\mu^-\nu_{\ell}\bar{\nu}_{\ell})_{SM+Z'}$ signal), the approximations I-V are not faithful to the kinematics and we might get unphysical results, i.e. $|\Mc^{mod}_{\tau\tau}|^2 < 0 $. To get an idea of how negative matrix element squared can arise, let us look at the conclusion of assumption IV: $ p_T^{\tau} = p_T^{ \nu s_i}+ p_T^{\mu_i}$, which means $p_T^{\nu s}$ is calculated based on $p_T^\tau$. We can determine $p_T^\tau$ using conservation of energy and momentum and a combination of assumptions. We get 
 \beq
 (p_T^{\tau})^{\rm approx} = \sqrt{ \frac{ M_Z^2}{2(1+ \cosh (\Delta \eta(\mu^+, \mu^-)))} - m_\tau^2}.
 \eeq
 Consequently, the $p_T$ of the vector sum of the two neutrinos coming from a tau can be deduced from these approximations: $p_T^{\nu s_i} = (p_T^{\tau})^{\rm approx} - p_T^{\mu_i}$. These approximations were reasonable in the framework of $ \left. \tau^+ \tau^-\right|_{\text{dimuon}}$. However, for $(pp \to \mu^+\mu^-\nu_{\ell}\bar{\nu}_{\ell})$ we can have $(p_T^{\tau})^{\rm approx} < p_T^{\mu_i}$ and therefore a negative (unphysical) magnitude for the transverse transverse momentum, which may lead to negative $|\Mc^{mod}_{\tau \tau}|^2$. The weights (area normalized) of $|\Mc^{mod}_{\tau \tau}|^2$ for MC generaed $(pp \to \tau^+\tau^- \to \mu^+\mu^-\met)$, $(pp \to \mu^+\mu^-\nu_{\ell}\bar{\nu}_{\ell})_{SM}$ and $(pp \to \mu^+\mu^-\nu_{\ell}\bar{\nu}_{\ell})_{SM+Z'}$ are shown below in Fig.~\ref{fig:tataME}. As expected, the $ \left.\tau^+ \tau^- \right|_{\text{dimuon}}$ distribution is more localized and all of its events have $|\Mc^{mod}_{\tau \tau}|^2 > 0$. Therefore, if we restrict ourselves to events with $|\Mc^{mod}_{\tau \tau}|^2 < 0$, we can safely assume that $ \tau \tau$ background is negligible. We have generated 10 million $ \left. \tau^+ \tau^-\right|_{\text{dimuon}}$ events, and 1 million events for each of the $(pp \to \mu^+\mu^-\nu_{\ell}\bar{\nu}_{\ell})_{SM}$ and $(pp \to \mu^+\mu^-\nu_{\ell}\bar{\nu}_{\ell})_{SM+Z'}$ processes, to make sure we have captured the tail of the $ \left. \tau^+ \tau^-\right|_{\text{dimuon}}$ distribution correctly.

\begin{figure}[h!]
\centering 
\includegraphics[width=0.65\textwidth, height=0.4\textheight]{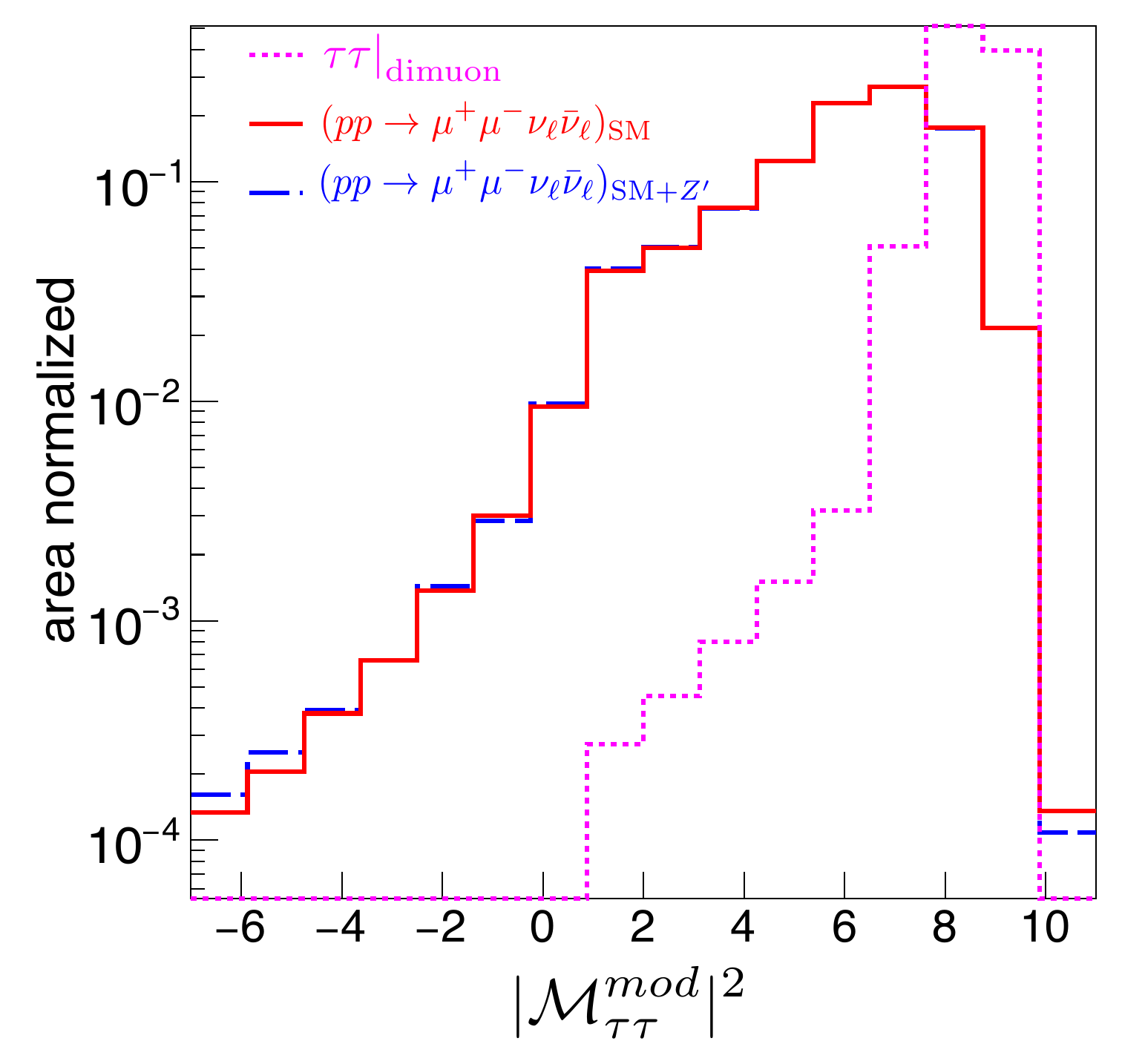}
\caption{The weights $|\Mc^{mod}_{\tau\tau}|^2$ for various MC sample events. The dotted purple line shows the weights for MC generated $ \left. \tau^+ \tau^-\right|_{\text{dimuon}}$ (10 million generated events). The solid red line shows the MC $(pp \to \mu^+\mu^-\nu_{\ell}\bar{\nu}_{\ell})$ background (1 million events), and the dashed blue line is MC $(pp \to \mu^+\mu^-\nu_{\ell}\bar{\nu}_{\ell})$ including $Z'$ (1 million events). Because the assumptions were chosen based on $ \left.\tau^+ \tau^- \right|_{\text{dimuon}}$ topology and were not reasonable in other processes, we have $|\Mc^{mod}_{\tau \tau}|^2<0 $ for part of the distribution of other processes. 
}
\label{fig:tataME}
\end{figure}

\begin{figure}[h!]
\centering 
\includegraphics[width=0.6\textwidth, height=0.39\textheight]{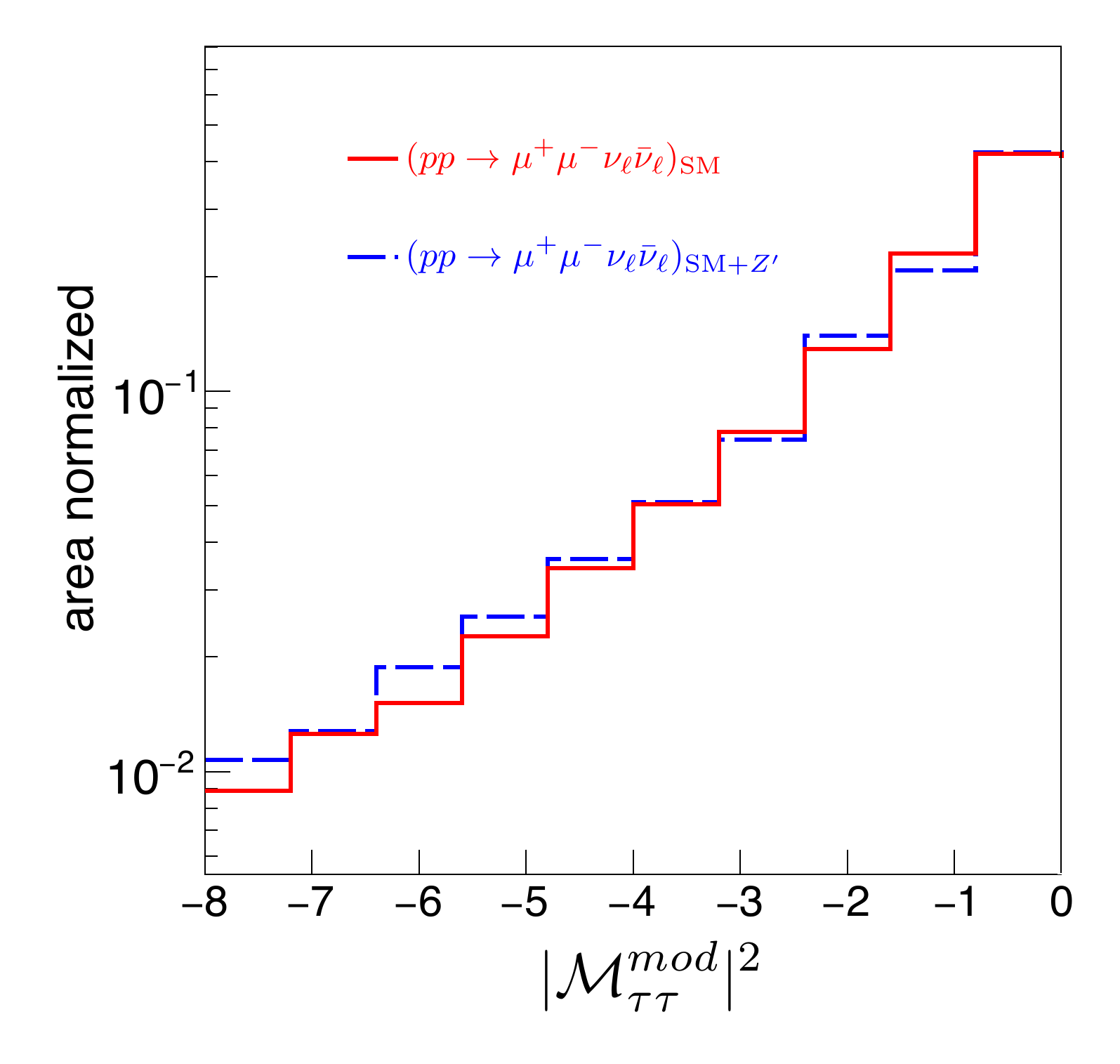}
\caption{ The weights $|\Mc^{mod}_{\tau\tau}|^2$ for $(pp \to \mu^+\mu^-\nu_{\ell}\bar{\nu}_{\ell})$ background (SM only, solid line) and signal (SM + $Z'$, dashed line) in the negative $|\Mc^{mod}_{\tau\tau}|^2$ region. 
}
\label{fig:tataME2}
\end{figure}

Inspecting Fig.~\ref{fig:tataME}, we can see that the signal $(pp \to \mu^+\mu^-\nu_{\ell}\bar{\nu}_{\ell})_{SM+Z'}$ and background $(pp \to \mu^+\mu^-\nu_{\ell}\bar{\nu}_{\ell})_{SM}$ have slightly different weights according to $|\Mc^{mod}_{\tau \tau}|^2$. This is not completely surprising because there are contributions from more diagrams in the signal events. The difference is most significant at large, negative $|\Mc^{mod}_{\tau \tau}|^2$; in particular, $(pp \to \mu^+\mu^-\nu_{\ell}\bar{\nu}_{\ell})$ including $Z'$ intermediate states populates $|\Mc^{mod}_{\tau \tau}|^2 < -4 $ more than when the $Z'$ is excluded. This $|\Mc^{mod}_{\tau \tau}|^2 $ region is shown in greater detail in Figure \ref{fig:tataME2}.

 To comprehend why $(pp \to \mu^+\mu^-\nu_{\ell}\bar{\nu}_{\ell})_{SM+Z'}$ prefers to be in the region $|\Mc^{mod}_{\tau \tau}|^2 < -4 $, let us look once again at squared matrix element of the signal 
 \begin{align*}
 |\Mc_{\rm signal}|^2 = |\Mc_{SM} + \Mc_{Z'}|^2 &= |\Mc_{Z'}|^2 + 2 | \Mc^*_{SM} \Mc_{Z'}| + |\Mc_{SM} |^2\\
 & \supset |\Mc_{Z', \text{ on-shell}}|^2 + 2 | \Mc^*_{WW} \Mc_{Z'}|,
 \end{align*}
where $\Mc_{WW}$ refers to the portion of the SM matrix element that involves the $WW$ contribution. Obviously, the departure of the signal from the SM background is most ideal for discrimination when $Z'$ is on-shell\footnote{The greatest contribution to the signal (only considering the $Z'$ contribution) is when on-shell $Z'$ comes from an on-shell $Z$. On the other hand, because both $ \left.\tau^+ \tau^- \right|_{\text{dimuon}}$ background and signal have $ \sqrt{\hat s} \lesssim M_{Z}$, a significant fraction of the signal (only the $Z'$ contribution) removed when removing $ \tau \tau$ background. This is inevitable, and the same challenge was faced with the cut and count method described in ~\cite{Elahi:2015vzh}, when $M_T(\mu \mu, \met) < 100 \ \gev$ was imposed to remove $ \left.\tau^+ \tau^- \right|_{\text{dimuon}}$. Therefore, it is really important to use the interference to look for our signal. }. That is because 1) the $|\Mc|^2$ is suppressed by only two powers of $\epsilon$ -- only one powers of $\epsilon$ at the production vertex of $Z'$ (amplitude level) and no $\epsilon$ suppression at the decay vertex, and 2) the topology of process with an on-shell $Z'$ mediator is different from the SM background, and so with some careful cuts we can make the SM background small. The next most important contribution of the signal is in the $Z'-SM$ interference, also suppressed by only two powers of $\epsilon$. The interference term is significant when either the SM piece is sizable or when the portion with $Z'$ contribution is big. Each option requires different kinematics; large $Z'$ contribution means the invariant mass of the neutrinos is small (or equivalently angular separation between the neutrinos is small), while large contribution of the SM could be when some of the intermediate states are produced on resonance. The only SM background that can have on-shell resonances and yet have other kinematics consistent with a (nearly) on-shell $Z'$ is the t-channel $W^+W^-$ background. We suspect the excess in the signal in the region of $|\Mc^{mod}_{\tau \tau}|^2< -4$ is due to the interference of $Z'$ piece with the $WW$ contribution. 

\begin{figure}[h!]
\centering 
\includegraphics[width=0.85\textwidth, height=0.28\textheight]{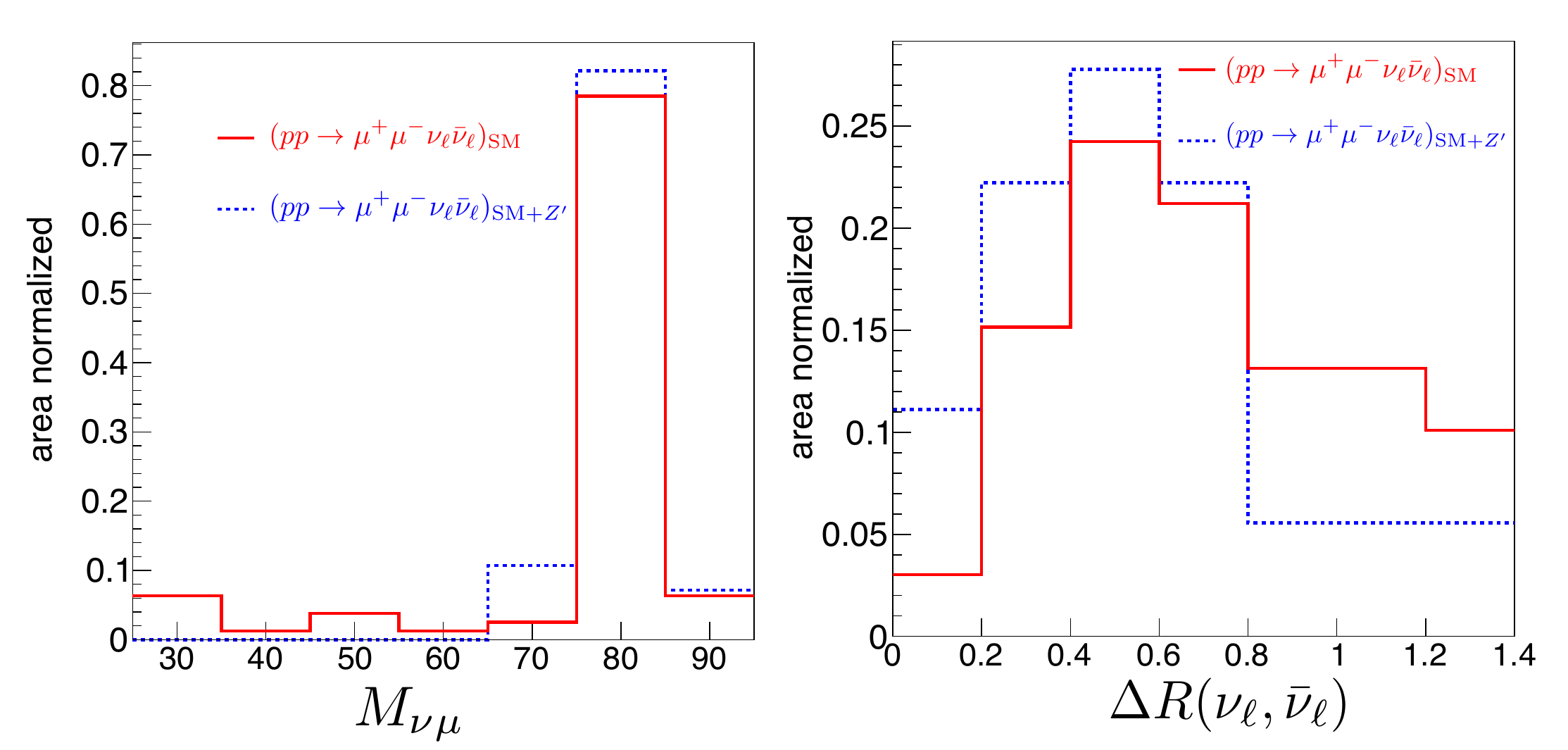}
\caption{ The left plot represents the invariant mass of a muon and the associated neutrino $ M_{\nu \mu}$, and the right plot is the separation between the two neutrinos $\Delta R(\nu_\ell, \bar \nu_\ell)$. These plots show the distribution of the events after requiring $|\Mc_{\tau \tau}^{mod}|^2 < -4 $, and demonstrate the signal events belong to on-shell $W$ production with $Z'$ (near) on-shell as well.   }
\label{fig:aftertata}
\end{figure}

 To check this intuition, we study the distributions of the events in the invariant mass of a muon and the associated neutrino (i.e. $M_{\nu_\ell \, \mu^+}$ or $M_{\bar \nu_\ell \, \mu^-}$), and the separation between the two neutrinos $(\Delta R (\nu_\ell, \bar \nu_\ell))$ after requiring $|\Mc^{mod}_{\tau \tau}|^2 < -4$ in Fig.~\ref{fig:aftertata}. These are not kinematic variables that we could actually measure, as neutrinos are not observed at the detector. However, studying them can help us understand the behavior of the simulated events for different regions of $|\Mc^{mod}_{\tau \tau}|^2$. 

 We can see from Fig.~\ref{fig:aftertata} that in the region of $|\Mc_{\tau \tau}^{mod}|^2 < -4 $, all of the signal events have $M_{\mu\nu} \sim M_W$, and have a small separation between the two neutrinos. This is perfectly consistent with what we expected from the interference term; the effect of the interference is enhanced when $W$s are near resonance, and other kinematic distributions are more consistent with $Z'$ signal. Thereby, we can be confident that the excess at  $|\Mc_{\tau \tau}^{mod}|^2 < -4 $ is due to the interference between the signal and the $W^+W^-$ background. 

With the basic cuts, $\met > 20 \ \gev$, and modified $|\Mc_{\tau \tau}|^2 < -4 $, we get the cross section of the background $\sigma(pp \to \mu^+\mu^-\nu_{\ell}\bar{\nu}_{\ell})_{SM}$ of $80\pm 3$ ab, while 
$\sigma(pp \to \mu^+\mu^-\nu_{\ell}\bar{\nu}_{\ell})_{SM+Z'}$ is $133\pm 4$ ab. The uncertainties on the cross sections are derived based on the number of events in our simulation that survive the imposed cuts. Let us define: 
 \begin{align*}
 S &\equiv \text{Luminosity} \times \left(\sigma( p p \to \mu^+ \mu^- \nu_\ell \bar{\nu_\ell} )_{\rm SM+Z'} - \sigma( p p \to \mu^+ \mu^- \nu_\ell \bar{\nu_\ell} )_{\rm SM}\right) \\
 B& \equiv \text{Luminosity} \times \left(\sigma( p p \to \mu^+ \mu^- \nu_\ell \bar{\nu_\ell} )_{\rm SM} + \sigma (p p \to \tau^+ \tau^- \to \mu^+ \mu^- \nu_\mu \bar{\nu_\mu} \nu_\tau \bar{\nu_\tau})_{SM}\right). 
 \end{align*}
 If we use the significance $S/\sqrt{B}$ as a test statistic and assume that $p p \to \mu^+ \mu^- + \text{jets}$ (which was our reducible background) is zero \footnote{This assumption is backed up by a MC study of 500K  $p p \to \mu^+\mu^-$ events of generated at the detector level ({\tt PGS}~\cite{PGS}) with the default smearing algorithm. Requiring  events pass the dilepton trigger and contain no jets with $p_T^{j} > 20 \, \gev$ and photons with $p_T^\gamma > 10 \, \gev$, the cross section was 111 \text{pb}. After imposing $\met = p_T(\text{dimuon}) > 20 \, \gev$, the cross section drops to $0.22 \, \text{pb}$. With the further requirement of $ |\Mc_{\tau \tau}^{mod}|^2 < -4$, we find the $p p \to  \mu^+\mu^-$ background can be removed completely.} after imposing the \MET cut and $|\Mc_{\tau \tau}^{mod}|^2 < -4 $, we find that after $300 \ \rm fb^{-1}$ luminosity, we get $S/\sqrt{B} \geq 3$. 
 
\begin{figure}[h!]
\centering 
\includegraphics[width=0.68\textwidth, height=0.35\textheight]{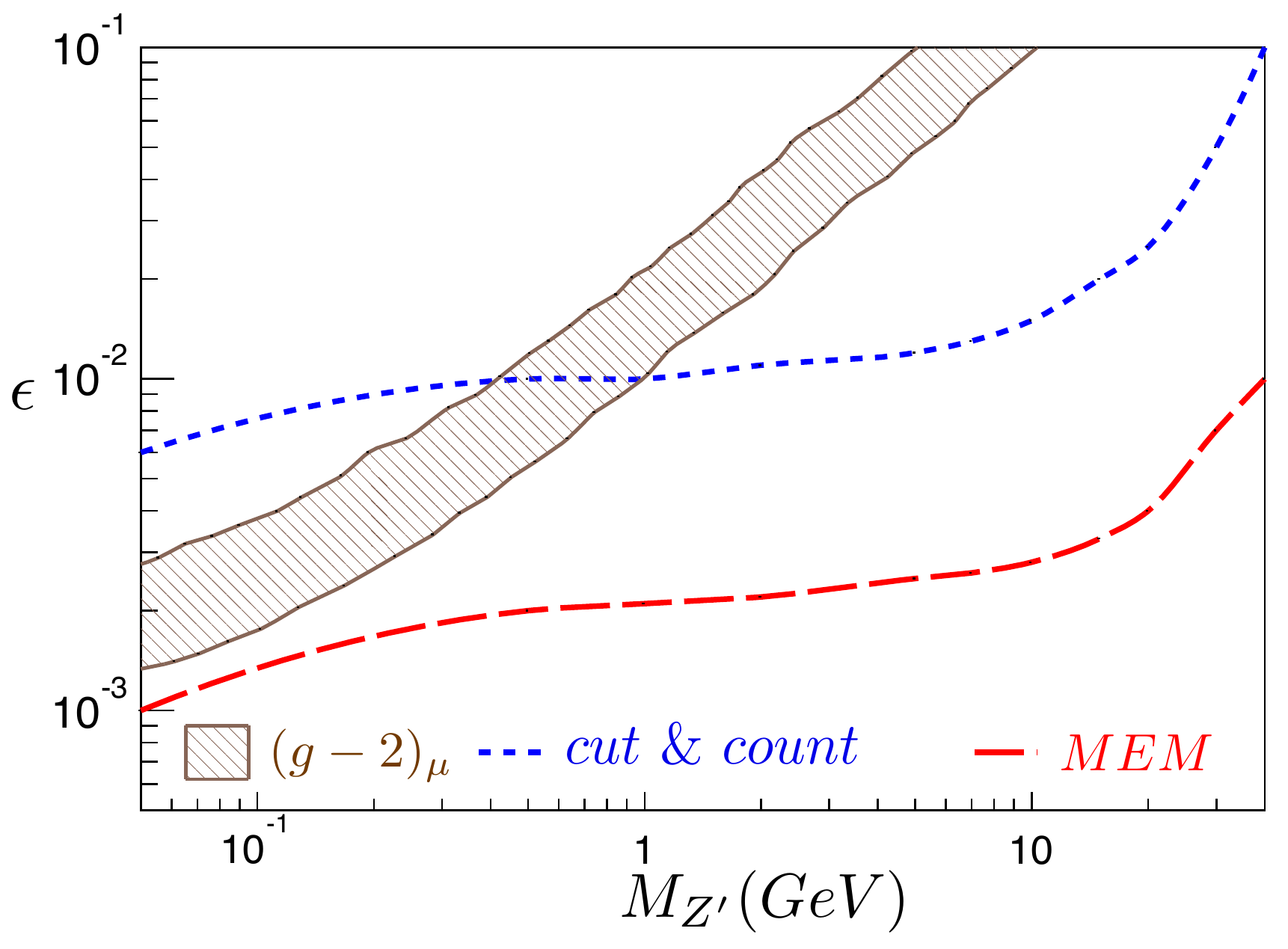}
\caption{ The new bounds according to our study using MEM compared with the cut and count method~\cite{Elahi:2015vzh}. The $(g-2)_\mu$ band is also shown in brown. For $M_{Z'} > 2m_\mu$, the MEM bound depends on whether we know $M_{Z'}$ or not. The red line shown here is our best bound. For $M_{Z'} < 2m_\mu$, our analysis is independent of $M_{Z'}$. The bounds from MEM and cut and count method are with luminosity of $300 \ \rm fb^{-1}$ and are up to $3 \ \sigma$.}
\label{fig:eps-mzp2}
\end{figure}

 Our reach in the region of the parameter space using MEM in contrast with the cut-and-count method and the $(g-2)_\mu$ band is shown in Fig.~\ref{fig:eps-mzp2}. The MEM and cut-and-count method bounds are based on the benchmark point of $M_{Z'}^{\text{gen}} = 0.05 \, \gev$ and $\epsilon^{\text{gen}} =0.001$. We see that with MEM, our sensitivity improves by a factor of 5-10 compared to the cut-and-count method, and we can explore a greater region of parameter space including the $(g-2)_\mu$ band. 
 
\section{ Discusssion}
\label{sec:discuss}

In this paper, we applied the Matrix Element method (MEM) to the $L_\mu-L_\tau$ model and concluded that our sensitivity improves by up to an order of magnitude compared with the cut-and-count method. The MEM uses the matrix element of a process to distinguish signal from background, and it has proven to be a powerful tool in several Standard Model (SM) measurements (e.g, top quark mass) and searches (e.g, Higgs to four lepton and electroweak single top production). However, it has not been extensively applied to beyond the SM (BSM) searches. The two main hurdles are that the MEM needs the physical parameters of the model and the four momenta of all initial and final states as inputs. In the quest of finding BSM signatures, we do not know the value of the new model parameters and the processes often contain unknown momenta in the form of missing energy. 

To investigate how we can combat these difficulties, as a first small step, we applied the MEM to the $L_\mu-L_\tau$ model. This model, being one of the simplest extensions of the SM, is already very well motivated because it can explain some of the current observational anomalies in the $(g-2)_\mu$ and B decays. In this model, there is a $Z'$ that couples to only second and third generation leptons at tree level. As a result, any tree-level process at the LHC involving $Z'$ has to include four leptons of second or third generation. We considered two cases, $4\mu$ and $2\mu+\met$. Both of these processes have a large number of kinematic observables, making them ideal test grounds for the MEM approach. The new parameters introduced by the $L_\mu-L_\tau$ model are the $Z'$ mass $M_{Z'}$ and coupling (parameterized by $\epsilon$: $g_{Z'} = \epsilon g'$).

In the mass range $ 2m_\mu < M_{Z'} < M_{Z} $, we looked at the process $pp \to Z \to 4 \mu$. This channel is clean and well-understood, and the presence of $Z \to 4 e$ and $Z \to 2e\ 2\mu$ control samples can be used to mitigate systematic uncertainties. Therefore, we can be sensitive to percent-level deviations. As the matrix element of the signal depends on $M_{Z'}$ and $\epsilon$, we first discussed how we can find values of these parameters that best separate signal from background. The optimal value of $M_{Z'}$ can be found by maximizing likelihood ratio with respect to $M_{Z'}$, depending on the strength of the $Z'$ coupling. However, the likelihood ratio function increases monotonically as a function of $\epsilon$ regardless of whether events belong to the signal sample or the background sample. Hence, we had to deviate from the conventional MEM and look for the most optimal analysis value of $\epsilon$ by studying the distribution of the signal and background MC generated events as a function of likelihood ratio for various fixed values of $\epsilon$. In the best case scenario, $L_\mu-L_\tau$ model can be explored up to 3$\sigma$ for $ \epsilon \gtrsim 0.002$ for $ 2m_\mu < M_{Z'} < 20 \ \gev $ and $ \epsilon \gtrsim 0.005$ for $ 20 < M_{Z'} < 40 \ \gev $ with luminosity of $300 \ \rm fb^{-1}$ at the LHC, which is about an order of magnitude improvement compared with the cut-and-count method. 

For lighter $Z'$, we studied the process $p p \to 2 \mu \met$. Due to the presence of missing energy, this channel is not as clean as the all muonic final state and is afflicted by several backgrounds. One significant background is $pp \to \tau^+ \tau^-$ with the taus decaying to muons, which has a cross section that is orders of magnitude greater than signal. To be sensitive to the signal, we first work towards eliminating the $\left.\tau^+ \tau^-\right|_{\text{dimuon}}$. We use the squared matrix element of $ \left.\tau^+ \tau^- \right|_{\text{dimuon}}$ for this task. This is a departure from the canonical MEM, as we weight the events by the squared matrix element of only one process, whereas in MEM we usually weight the events by the ratio of the squared matrix element of the signal processes over the background ones. This alternative approach has two main benefits: 1) given that we have missing energy in the process, calculating the squared matrix elements is challenging; Hence, focusing our attention to only one can save us time and computational power. 2) Furthermore, this approach is independent of the model parameters and can be used for any BSM physics with this signature at the LHC. 

Due to the presence of 4 sources of missing energy in the $\left.\tau^+ \tau^-\right|_{\text{dimuon}}$ process, we have 10 unconstrained momenta in this channel. Traditionally, one would proceed by integrating over the unknown momenta. In this paper, we instead showed how the unknown momenta could be estimated by exploiting the topology of the $ \left.\tau^+ \tau^- \right|_{\text{dimuon}}$. Without the need for any integrations, we calculate the (modified) squared matrix element of $ \left.\tau^+ \tau^- \right|_{\text{dimuon}}$ ($|\Mc_{\tau \tau}^{mod}|^2$) and use the resulting weight as a discriminant. With a judicious cut on  $|\Mc_{\tau \tau}^{mod}|^2$, we find the $ \left.\tau^+ \tau^- \right|_{\text{dimuon}}$ background can be completely eliminated. We then observe that the signal reacts differently to the $|\Mc_{\tau \tau}^{mod}|^2$ compared to other (non-$tau$) SM backgrounds, and trace the difference to interference between $Z'$ contributions to the amplitude and contributions containing two on-shell $W$'s.  As a result, we can differentiate the signal from {\em all} SM backgrounds using $|\Mc_{\tau \tau}^{mod}|^2$ alone.  With this method, we find we can reach to 3$\sigma$ up to $\epsilon \sim 0.001$ for $M_{Z'} < 2m_{\mu}$ assuming an integrated luminosity of $300 \ \rm fb^{-1}$, covering the $(g-2)_\mu$ band. This result may be improved if we relax some of the kinematic assumptions and instead integrate over a subset of the unconstrained momentum, something which may be worth investigating in the future. 

Because we did not use the squared matrix element of the signal in our analysis of $p p \to 2 \mu \met$, our procedure can be applied to any BSM searches with leptons and missing energy in the final states. Scenarios with leptons and missing energy are particularly well-motivated in many dark matter and dark photon searches at the LHC~\cite{Gramstad:2013loc, Aad:2012pxa, Aad:2014vma, Baek:2013fsa, Kruker:2013joa, Wittkowski:2013fua, Cheng:2013hna, Teroerde:2016dtv,Altmannshofer:2016jzy}. More generally, we argue that processes with \MET that have a specific topology can benefit from MEM, while not suffering from its computational challenges. 

In conclusion, in this paper we provide a working example where, after approximating unknown momenta using the topology of the process, the MEM yields a superior sensitivity compared to the cut-and-count method without having to integrate. Consequently, even if applying the canonical MEM to BSM searches has obstacles and appears to be computationally challenging, we may be able to \emph{modify} MEM with reasonable assumptions to ease the computational difficulty and yet gain a better sensitivity than the cut-and-count method. 

\section*{Acknowledgments}
\label{sec:ack}

We thank Nirmal Raj for his valuable suggestions on the draft and Joe Bramante, Rodolfo Capdevilla, Carlos Alvarado, and Antonio Delgado for useful discussions. This work was partially supported by the National Science Foundation under Grants No. PHY-1417118 and No. PHY-1520966.

\appendix
\section{The Modified Squared Matrix Element of $\left.\tau^+ \tau^-\right|_{\rm dimuon}$ Process ($|\Mc^{mod}_{\tau \tau}|^2$)}
\label{app:tataME}
In this appendix, we detail the calculation of $|\Mc^{mod}_{\tau \tau}|^2$. The first step is to define the four-vector of $\nu s$ (the vector sum of the two neutrinos coming from each tau) using the momenta of muons in the framework of assumptions discussed in Section~\ref{sec:twomu}, keeping in mind the $\nu s$ are not massless:
$$p^{\nu s_i} = \left(\sqrt{\left(p_T^{\nu s_i}\right)^2+ m_{\nu s_i}^2} \cosh \eta_{\nu s_i}, p_T^{\nu s_i} \cos \phi_{\nu s_i}, p_T^{\nu s_i} \sin \phi_{\nu s_i}, \sqrt{\left(p_T^{\nu s_i}\right)^2+ m_{\nu s_i}^2} \sinh \eta_{\nu s_i}\right),$$
where the subscript $i$ is defined such that $ \tau_i \to \mu_i \nu s_i$. In the following, we will define $p_T^{\nu s_i}, \, m_{\nu s_i}, \, \eta_{\nu s_i}, $ and $ \phi_{\nu s_i}$ in terms of known or measurable parameters:
\begin{align*}
p_T^{\nu s_i} &= \sqrt{ \frac{ M_Z^2}{2(1+ \cosh (\Delta \eta(\mu^+, \mu^-)))} - m_\tau^2} - p_T^{\mu_i}\\
m_{\nu s_i} & = \sqrt{ m_\tau^2 + 2 (p_T^{\mu_i})^2 - 2 \sqrt{ (p_T^{\mu_i})^2 (m_\tau^2 + (p_T^{\mu_i})^2+ (p_T^{\nu s_i})^2)}}\\
\eta_{\nu s_i} & = \eta_{\mu_i}\\
\phi_{\nu s_i} & = \phi_{\mu_i}+ \epsilon_\phi^i, \, \text{ where} \, \, \epsilon_\phi^i \ll 1,
\end{align*}
where we have ignored the muon mass $(m_\mu= 0)$, and $ \epsilon_\phi$s are calculated from the conservation of momenta in the transverse plane. 

Furthermore, one of the assumptions (III) in Section~\ref{sec:twomu} is that the taus are produced from an on-shell $Z$. Therefore, in calculating $|\Mc^{mod}_{\tau \tau}|^2$, we will also assume the process is $ Z \to \tau^+ \tau^- \to \mu_1 \mu_2 \, \nu s_1 \nu s_2$, where $Z$ is simply the vector sum of the final state products, shown by $p^Z$. The modified squared matrix element of $\left.\tau^+ \tau^-\right|_{\rm dimuon}$ is 
\begin{align*}
|\Mc^{mod}_{\tau \tau}|^2 &= \frac{m_\tau^2}{(m_\tau^2 - m_{\nu s_1}^2)^{2}(m_\tau^2 - m_{\nu s_2}^2)^{2}} \left(\frac{ 8 m_{\nu s_1} m_{\nu s_2}(p^Z\cdot p^{\mu_1}) (p^Z\cdot p^{\mu_2})}{m_\tau^2 M_Z^2} + \frac{ 4 m_{\nu s_1}m_{\nu s_2} (p^{\mu_1}\cdot p^{\mu_2})}{m_\tau^2} \right.\\
&- \frac{8 m_{\nu s_1} [(p^Z\cdot p^{\mu_1}) (p^Z\cdot p^{\mu_2}) + 2 (p^Z\cdot p^{\nu s_1}) (p^Z\cdot p^{\mu_2})]}{m_\tau M_{Z}^2} - \frac{ 2 m_{\nu s_1} [(p^{\mu_1}\cdot p^{\mu_2}) + 2 (p^{\nu s_1}\cdot p^{\mu_2})]}{m_\tau} \\
&-\frac{ 4 m_{\nu s_2}[(p^Z\cdot p^{\mu_1})(p^Z\cdot p^{\mu_2})+2(p^Z\cdot p^{\mu_1})(p^Z\cdot p^{\nu s_2})]}{m_\tau M_{Z}^2} - \frac{ 2 m_{\nu s_2} [ (p^{\mu_1}\cdot p^{\mu_2})+ 2(p^{\mu_1}\cdot p^{\nu s_2}) ]}{m_\tau} \\
&+ \frac{1}{M_{Z}^2} [ 4 (p^Z\cdot p^{\mu_1}) (p^Z\cdot p^{\nu s_2}) + 2 (p^Z\cdot p^{\mu_1})(p^Z\cdot p^{\mu_2})+ 8 (p^Z\cdot p^{\nu s_1}) (p^Z\cdot p^{\nu s_2})]\\
&\left. \frac{4}{M_{Z}^2} (p^Z\cdot p^{\nu s_1})(p^Z\cdot p^{\mu_2}) + 2 (p^{\mu_1} \cdot p^{\nu s_2})+ (p^{\mu_1}\cdot p^{\mu_2})+4 (p^{\nu s_1}\cdot p^{\nu s_2})+ 2 (p^{\nu s_1}\cdot p^{\mu_2})\right)
\end{align*} 

The numerical coefficient in front of $|\Mc^{mod}_{\tau \tau}|^2$ is irrelevant and thus can be ignored. This is because it does not matter whether the plots are with respect to $|\Mc^{mod}_{\tau \tau}|^2$ or $16 \pi |\Mc^{mod}_{\tau \tau}|^2$.

\bibliography{mutau}
\bibliographystyle{JHEP}

\end{document}